\documentclass{article}

\usepackage{PRIMEarxiv}

\usepackage[utf8]{inputenc}
\usepackage[T1]{fontenc} 
\usepackage{hyperref} 
\usepackage{url} 
 
\usepackage{amsfonts} 
\usepackage{nicefrac}
\usepackage{microtype}
\usepackage{lipsum}
\usepackage{fancyhdr}  
\usepackage{graphicx} 
\graphicspath{{media/}}
\usepackage{lmodern}
\usepackage{babel} 
\usepackage{graphicx}
\usepackage{enumerate}
\usepackage{amssymb}
\usepackage{lipsum}  
\usepackage{amsmath}
\usepackage{amsthm}
\usepackage{dsfont}
\usepackage{bm}
\usepackage{todonotes}
\usepackage{natbib}
\usepackage{hyperref}
\usepackage[capitalize,nameinlink]{cleveref}
\usepackage{lipsum}  
\usepackage{xcolor}
\usepackage{float}
\usepackage{booktabs}

\usepackage{bm}
\usepackage{dsfont}

\newcommand{\ind}{\mathds{1}}
\newcommand{\R}{\mathds{R}}

\newcommand{\Xcal}{\mathcal{X}}

\newcommand{\eps}{\varepsilon}

\providecommand{\Pr}{}
\renewcommand{\Pr}{\mathbb{P}}

\newcommand{\wt}[1]{\widetilde{#1}}

\makeatletter
\def\blfootnote{\gdef\@thefnmark{}\@footnotetext}
\makeatother

\definecolor{saale}{HTML}{450D54}
\definecolor{iller}{HTML}{B9534C}
\definecolor{ziller}{HTML}{22908C}

\graphicspath{{Fig/}}

\newtheorem{theorem}{Theorem}
\newtheorem{proposition}[theorem]{Proposition}%

\title{Towards more realistic climate model outputs: \\ A multivariate bias correction based on zero-inflated vine copulas}

\author{
  Henri Funk \\
  Department of Geography \\
  Statistical Consulting Unit StaBLab \\
  Munich Center for Machine Learning, MCML\\
  LMU Munich \\
  Munich\\
  \texttt{H.Funk@lmu.de} \\
   \And
  Ralf Ludwig \\
  Department of Geography \\
  LMU Munich \\
  Munich\\
  \AND
  Helmut Küchenhoff \\
  Statistical Consulting Unit StaBLab \\
  Munich Center for Machine Learning, MCML\\
  LMU Munich \\
  Munich\\
  \And
  Thomas Nagler \\
  Department of Statistics \\
  Munich Center for Machine Learning, MCML\\
  LMU Munich \\ 
  Munich
}

\begin{document}
\maketitle

\begin{abstract}
Climate model large ensembles are an essential research tool for analysing and quantifying natural climate variability and providing robust information for rare extreme events.
The models simulated representations of reality are susceptible to bias due to incomplete understanding of physical processes.
This paper aims to correct the bias of five climate variables from the CRCM5 Large Ensemble over Central Europe at a 3-hourly temporal resolution.
At this high temporal resolution, two variables, precipitation and radiation, exhibit a high share of zero inflation.  
We propose a novel bias-correction method, VBC (Vine copula bias correction), that models and transfers multivariate dependence structures for zero-inflated margins in the data from its error-prone model domain to a reference domain. 
VBC estimates the model and reference distribution using vine copulas and corrects the model distribution via (inverse) Rosenblatt transformation.
To deal with the variables' zero-inflated nature, we develop a new vine density decomposition that accommodates such variables and employs an adequately randomized version of the Rosenblatt transform.
This novel approach allows for more accurate modelling of multivariate zero-inflated climate data.
Compared with state-of-the-art correction methods, VBC is generally the best-performing correction and the most accurate method for correcting zero-inflated events.
\end{abstract}

% keywords can be removed

\section{Introduction}\label{sec:int}

\subsection{Motivation}

Physical climate simulations encompassing global and regional climate models (GCMs and RCMs) are fundamental to understanding the climate and projecting future climate scenarios.
In recent scientific analysis, climate simulations are bundled into model ensembles which can be used to quantify natural climate variability, as highlighted by \citet{vicente2021} and \citet{vonTrentini2019}.
In addition, large ensemble GCMs and RCMs are crucial for understanding and quantifying rare extreme events in historical, contemporary, and projected future climates.

\begin{figure}[H]
    \centering
    \includegraphics[width = 0.95\textwidth]{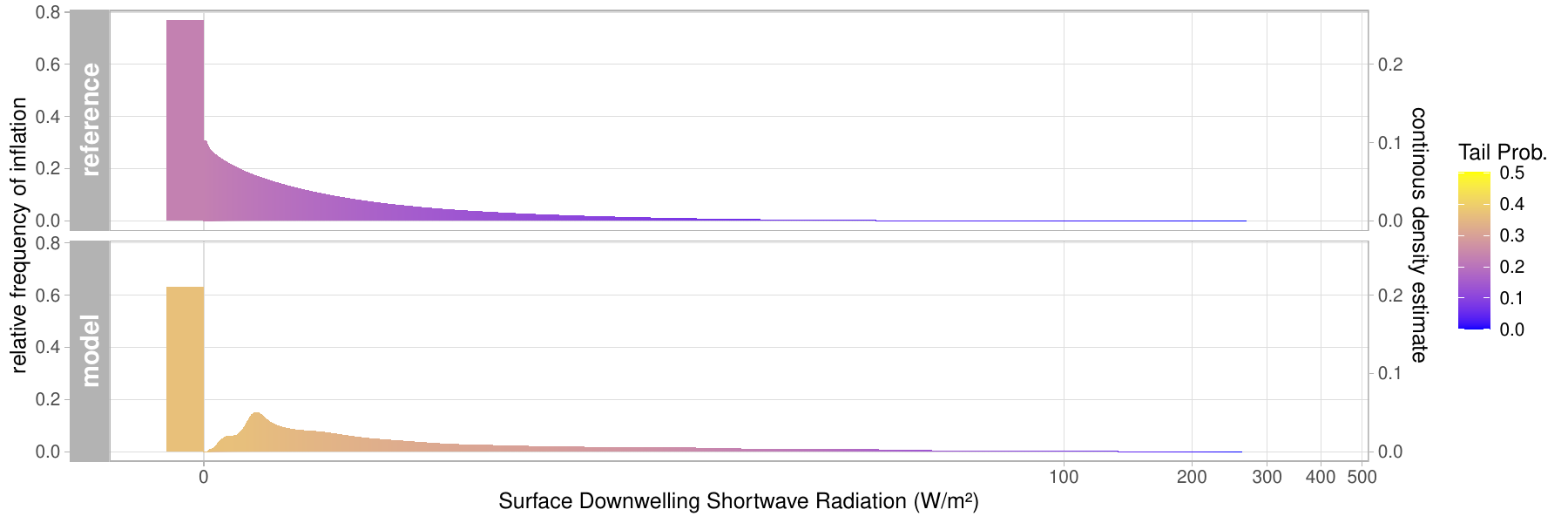}%\rule{438pt}{74pt}
    \caption{Zero-inflated distribution of 3-hourly averaged radiation in Iller-Kepmten in the Allgäuer High Alps (47°20'14.64N, 10°12'53.244E) in winter during nighttime.
             In the rows, the densities are faceted by the CRCM5-LE (model) data and the SDCLIREF v2 (reference) data.
             The bar indicates the relative inflation of the discrete, and the density curve represents the climate variable's continuous domain.
             For comparison, the colour gradient encodes the tail probabilities of radiation with yellow encoding the median and blue encoding the tails.}
    \label{fig:mVr}
\end{figure}

GCMs and RCMs estimate instances of meteorological reality.
Compared to the reference data, these climate models show discernible \textit{biases} due to simplified or incomplete atmospheric or physical processes or insufficient understanding of the global climate system \citep[e.g.]{chen2021}.
These biases are not covered by the internal variability of the ensemble members \citep{sorland2018, franccois2020}.
For instance, a pronounced hot bias signifies a model's tendency to overestimate observed temperatures, whereas a wet bias denotes an overprediction of precipitation levels.
However, the bias within these climate models usually transcends a mere additive shift but affects the entire multivariate distributional shape of the climate variables. 
\cref{fig:mVr} displays the distribution function of radiation for model and reference data.
The exemplified model distribution has a pronounced bias in the zero inflation by underestimating time steps without radiation.
The model shows an unsteady shape compared to the smoother reference in the lower continuous support of the distribution function between 0 and 250 $W/m^2$.
The heavy tail of the model distribution is overestimated, producing more instances of higher radiation during the night than the historical reference.
All the divergences between the model and the reference distribution function described above are considered biases.

Bias correction (BC), or adjustment, methods aim to rectify these errors in model data using information from historical data as a distributional reference while preserving the model's specific pattern.
Ideally, the reference data represent a recent historical climate period of about 30 years called the calibration period \citep{reiter2016}.
Climate model data outside the calibration period can also corrected by the reference data and subsequently projected to their native temporal domain, e.g. by Delta Mapping \citep{cannon2015}.
Delta Mapping accounts for the discrepancy in climatic conditions between the calibration period and past or future data.

Bias correction is an advisable step before downscaling (regional) climate model simulations to higher resolutions, as it adjusts local inconsistencies in the climate model.
The corrected and downscaled meteorological inputs can be used directly for meteorological climate impact analysis \citep{chen2013} or to drive a hydrological model \cite[e.g.]{emami2018, fang2015}.
The results of the hydrological flow modelling serve as inputs to the quantification and analysis of the drivers of floods or low flow events \citep{soriano2019, hashino2007}. 
The reliability of the succeeding methods and analysis depends on the quality of corrected data an hence on the BC method used \citep[e.g.]{ahmed2013, chen2013, willkofer2018}.

\subsection{Previous Work}

The quantile delta mapping technique (UBC) \citep{cannon2015} calculates the cumulative density function for one climate variable from the model and the reference data. 
This allows for a rank-based data transfer from the model's native domain to the climate variable's reference domain.
As a result, the scale of the model data is adapted to the scale of the reference, making each climate variable more realistic.
Implicitly, the inherent rank structure specific to the models' climate variable is preserved.
UBC operates at an univariate level and should only be considered if one climate variable is to be corrected at a time and multivariate physical dependencies are of minor interest. 

Climate variables often exhibit specific correlation structures among each other, e.g., due to physical connections.
The correlation structures can result from locally diverse climates that are affected by altitude, vegetation, or other climate factors.
For example, dewpoint temperature is a result of temperature and relative humidity and is, therefore, strongly connected to temperature.
Relative humidity might vary depending on the vegetation coverage.
\citet{guo2020} has drawn out the benefits of a multivariate bias correction method for hydrological modelling.
To account for joint dependencies in climate variables, the Multivariate Bias Correction algorithm (MBCn) \citep{cannon2018} generalises the univariate UBC to deal with multivariate distributions by implementing the n-pdft algorithm \citep{pitie2005}.
MBCn employs an iterative process that uses a random orthogonal rotation matrix to project the climate data onto a space where the variables are mutually independent and the bias can be corrected per variable by univariate quantile mapping (UBC) before the rotation is inverted.
Many other task-specific multivariate bias corrections exist, e.g. R2D2, a stochastic, rank-based, multivariate bias correction that is specifically developed for high-dimensional problems \citep{vrac2018} or a correction that uses optimal transport to transfer distributional characteristics \citep{robin2019}.
Note that, to the best of our knowledge, neither of those methods inherently accounts for zero inflation in the variables. 
MBCn implements a heuristic truncation at a small positive threshold to artificially reintroduce zero inflation after the correction.

However, many climate variables at high temporal resolution, like sub-daily precipitation and radiation, exhibit zero inflation. 
Therefore, the issue of modelling zero-inflated margins in climate data is a current scientific concern.
\citet{maity2019} propose a bivariate correction method using copulas designed to address zero inflation in precipitation.
This method adjusts the conditional distribution of the zero-inflated variable using a discrete segment within the correction process.

\subsection{Main contributions}

In this paper, we aim to correct the bias in a climate model large-ensemble with 3-hourly resolutions, partially zero-inflated and heavy-tailed climate variables, and physical dependencies between the climate variables.
Our main contribution is \textbf{Vine Copula Bias Correction for partially zero-inflated margins (VBC)}, a novel multivariate bias correction methodology for climate models anchored in vine copula theory.
The foundation of the technique is a generalization of vine copula theory and methods to accommodate variables that consist of a mixture of discrete and continuous components, such as the zero-inflated variable shown in \cref{fig:mVr}.
The main theoretical result, \cref{prop:vines}, generalizes the vine density decomposition formula from \citet{bedford2001, bedford2002} to such variables.
This is combined with a Rosenblatt transformation that allows the adjustment of simulated data from biased climate models to an appropriate reference distribution, both in terms of marginal distributions and dependence structure.
The transformation is randomized to accommodate non-continuous variables.
Although this was previously proposed by \citet{brockwell2007}, it is new in the context of vine copula models.
To project the climate data from the calibration into past or future climates, we adapt the Delta Mapping from \citet{cannon2015} for a more cautious projection of heavy-tailed variables.
In particular, our approach is designed for high interpretability, enabling control and assessment of the results.
Finally, we demonstrate the exceptional performance of our method by comparing the corrected data against other leading bias correction techniques in a real-data application.
The results are evaluated by Wasserstein distance and the Model Correction Inconsistency, a new measure that we introduce to account for the preservation of weather after multivariate bias correction.
The data sets involved in the application are published and free to download \citep{funk2024a}.
The bias correction method is implemented alongside visualization and evaluation tools in R and is available as a fully installable package with a reproducible vignette on GitHub \citep{funk2024b}.

\subsection{Outline}

The paper is organized as follows.
\cref{sec:data} introduces the relevant climate data set, the regional scope of interest, and the concerning climate variables.
The procedure of the bias correction in \cref{sec:bc} consists of three elementary steps of VBC: modelling, correction, and projecting the climate data. \cref{sec:vines} introduces the necessary background on vine copula models and gives the main theoretical result; details on the three steps of the VBC method and their implementation are given in \cref{sec:vbc}. The results of our data application are discussed in \cref{sec:res}.
Finally, \cref{sec:dis} elaborates on the potentials of VBC and points out possible enhancements.

\section{Data and Correction Approach}

\subsection{Climate Data} \label{sec:data}

\begin{figure}[tp]
\centering
    \includegraphics[scale = 0.8]{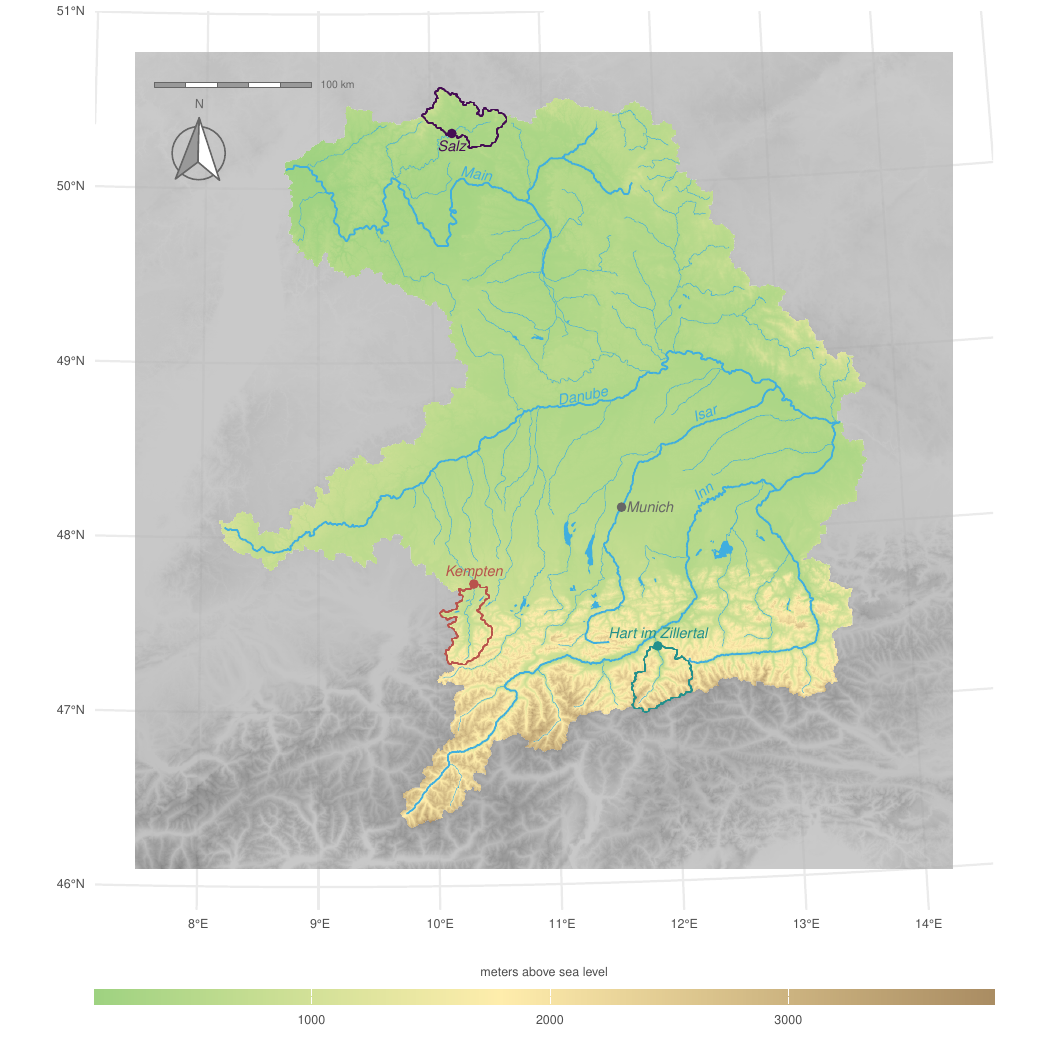}
    \caption{Bavaria and important hydrological neighbouring regions with three highlighted catchments: An alpine catchment following the Ziller river to Hart in Zillertal (cyan), a pre-alpine catchment at the Iller up until Kempten (red) and a Franconian catchment from the origin of the Fränkische Saale until Salz (purple). The biggest city in the region of interest, Munich, is indicated as a reference.\label{fig:hydbav}}% \caption{Figure caption}
    \label{fig:enter-label}
\end{figure}

The Canadian Regional Climate Model version 5 Large Ensemble (CRCM5-LE; \citep{leduc2019, martynov2013, separovic2013}) covers two domains across Europe and North America with a 12km resolution, covering the period from 1950 to 2099. 
This large ensemble comprises 50 transient members to distinguish between climate change signals and natural variability, each generated through dynamical downscaling.
CRCM5 is a SMILE (Single Model Initial-condition Large Ensemble) indicating that the 50 members are equally likely realizations of one GCM.
The downscaling employs the Canadian Earth System Model Large Ensemble (CANESM-LE), a GCM, as boundary conditions.
CRCM5 is employed under the Representative Concentration Pathway 8.5 (RCP8.5) and expects a total population of 12 billion humans on Earth, a mean temperature increase of 4.8°C and a decrease in precipitation between 50 and 75\% in southwest Europe by the end of 2100 compared to pre-industrial climate conditions \citep{ipcc2022}.

SDCLIREF v2 \citep{wood2024} is a sub-daily (3h), high-resolution (500m) climate reference data set for the extended domain over Bavaria in \cref{fig:hydbav} from 1990 until 2020.
The data set comprises five meteorological variables of hydrological importance to the water management in Bavaria outlined in \cref{tab:cv}.
Reference data are selected and processed historical climate data that are used in bias correction to correct and validate climate models.
Version 2, built upon SDCLIREF v1 \citep{wood2017}, employs the Method of Fragments \citep{srikanthan2006, westra2013} for temporal disaggregation.
Further, it adjusts wind-related measurement errors in precipitation in the SDCLIREF v1.

\begin{table}[H]
    \centering
    \begin{tabular}{clr}
        \toprule
        Index & Description & Units \\
        \midrule    
        $\texttt{d}$ & Near-Surface Dewpoint Temperature & °C \\
        $\texttt{p}$ & Precipitation & kg/m² \\
        $\texttt{r}$ & Surface Downwelling Shortwave Radiation & W/m² \\
        $\texttt{w}$ & Near-Surface Wind Speed & m/s \\
        $\texttt{t}$ & Near-Surface Air Temperature & °C \\
        \bottomrule
    \end{tabular}
    \caption{List of meteorological climate variables of hydrological importance. 
    All variables are available in a 3-hourly temporal resolution. \label{tab:cv}}
\end{table}

As detailed in \citet{leduc2019}, the CRCM5-LE model domain over Europe exhibits notable warm biases during winter months and cold biases in summer, alongside a tendency towards higher precipitation than observed.
In addition to these broad biases, the CRCM5-LE fails to model the meteorological patterns in singular grid cells, due to local climate conditions induced by vegetation, elevation, slope or exposition in the grid cell.
These local biases can distort the distribution of singular climate variables or their mutual physical dependencies.
Further, biases are time-dependent. 
A model bias of a variable during winter can be different compared to a bias during summer since many physical climate conditions change over time.
\cref{fig:mVr} exemplifies the complexity of a regional bias for Surface Downwelling Shortwave Radiation in the Allgäuer High Alps during winter nighttime.

\begin{table}[H]
    \centering
    \begin{tabular}{cllll}
        \toprule
        Notation & BC Usage & Data Set & Period & Years \\
        \midrule  
        $\mathcal{D}_{rc}$ & Reference data & SDCLIREF v2 &  Calibration& 1991 - 2010\\
        $\mathcal{D}_{rp}$ & Hold out reference & SDCLIREF v2 & Projection & 2011 - 2020\\
        $\mathcal{D}_{mc}$& Calibration data & CRCM5-LE & Calibration & 1991 - 2010\\
        $\mathcal{D}_{mp}$  & Projection data & CRCM5-LE & Projection & 2011 - 2020\\
        \bottomrule
    \end{tabular}
    \caption{Data sets used for correction and evaluation of biases in climate models. 
    1991 to 2010 are used to calibrate the bias correction method.
    The last 10 years from 2011 to 2020 are projection data.
    The Abbreviation $r$ and $m$ mark SDCLIREF v2 $r$eference and CRCM5 $m$odel data, respectively.
    $c$ and $p$ abbreviate $c$alibration (1991 -2010) and $p$rojection period (2011 - 2020), respectively.
    \label{tab:dat}}
\end{table}

The correction of the described characteristics of model biases is a necessary prerequisite in current climate research and applied methods like downscaling, hydrological modelling, hydrometeorological risk assessment and climate extremes analysis such as low flow and drought analysis \citep{haerter2011}.
Four types of data sets are derived from CRCM5 and SDCLIREF v2 and outlined in \cref{tab:dat} to correct for a bias and validate the correction.
$\mathcal{D}_{mp}$ marks a CRCM5-LE projection data set, i.e., the data set which should be corrected.
$\mathcal{D}_{rc}$ represents the SDCLIREF v2 reference data set.
In bias correction, $\mathcal{D}_{rc}$ contains the target distribution, i.e. the distribution that $\mathcal{D}_{mp}$ should be corrected to.
$\mathcal{D}_{mc}$ outlines the CRCM5-LE model data set during the calibration period.
Calibration data are used to preserve the ensemble's internal variability \citep{vaittinada2021} and to account for past and future changes in the model data, thus maintaining the RCP8.5 scenario.
$\mathcal{D}_{rp}$ denotes the SDCLIREF v2 data during projection period.
The data set is held out as a validation set to evaluate the correction results in \cref{sec:res}.

\subsection{Bias Correction}\label{sec:bc}

Our ultimate goal is to use information from the reference and calibration data sets to make the projected data as similar as possible to the unobserved future climate data. 
On the one hand, the climate model data should become more realistic.
On the other hand, the temporal course of the weather, expressed by each model member, should be preserved.
Technically, this means that after the correction, the climate data should be distributed like the reference climate data.
At the same time, the multivariate rank structure within each model member should be similar before and after the correction.
We propose a bias correction method consisting of the following three steps to satisfy these properties.

\begin{enumerate}[(a)]
    \item \textbf{Estimation step}: Model the multivariate distribution function of reference and model data sets.
\begin{equation} \label{eq:mvd}
    \begin{aligned}
        \mathcal{D}_{rc} &\sim \hat{F}_{rc} , \qquad \mathcal{D}_{mp} \sim \hat{F}_{mp}
    \end{aligned}
\end{equation}

For the data structure described in \cref{sec:data}, the estimation of the distribution functions $F_{rc}$ and $F_{mp}$ in \cref{eq:mvd} pose a problem of modelling a five-dimensional distribution with two zero-inflated margins, precipitation \texttt{p} and radiation \texttt{r}.
To model these dependencies, we develop a generalized vine decomposition of a multivariate density (\cref{prop:vines}) that applies to zero-inflated and other mixture variables in \cref{sec:zi}.
Details for the estimation of the multivariate distribution functions $\hat{F}_{rc}$ and $\hat{F}_{mp}$ and their partially zero-inflated margins are given in \cref{sec:vce}.\\

    \item \textbf{Correction step}: Transfer the distribution from the model to the reference domain. That is, construct new data 
\begin{equation*}
    \hat{\bm x}_{mc} := T(\bm x_{mp}) \quad \text{such that} \quad  \hat{\bm x}_{mc} \sim \hat{F}_{rc}.
\end{equation*}
For $T$, we propose chaining modified versions of the forward and inverse Rosenblatt transformation \citep{rosenblatt1952} that account for the inflation.
During the correction step of VBC, the forward Rosenblatt transform related to $\hat{F}_{mp}$ is used to transfer the data $\bm x_{mp}$ from its native climate model domain to a neutral uniform domain.
From here, we use the inverse Rosenblatt transform related to $\hat{F}_{rc}$ to transfer the uniform data to the more realistic domain $\hat{F}_{rc}$. 
%of SDCLIREF v2 in \cref{sec:rt}$$.
See \cref{sec:rt} for the details.

    \item \textbf{Projection step}:  Account for the discrepancy in climatic conditions between calibration and projection data. We do this by a mapping of the form
\begin{equation*}
    \hat{\bm x}_{mp} = \delta(\hat{\bm x}_{mc} | \bm x_{mp}, \bm x_{mc}).
\end{equation*}
The mapping $\delta$ estimates the discrepancy between the simulated projection data set $\bm x_{mp}$ and calibration data $\bm x_{mc}$, and adds this discrepancy to the corrected data. This step makes sure that the corrected data conforms with long-term climate developments (e.g., warming) reflected in the simulation model.
Our proposed solution in \cref{sec:dm} is a modification of the delta mapping proposed in \citet{cannon2015}. 
Details are given in \cref{sec:dm}.
In the application, the data are mapped from the calibration climate, from 1991 to 2010, onto the projection climate from 2011 to 2020.
\end{enumerate}

\section{Vine Copula Models for Mixture Distributions} \label{sec:vines}

\subsection{Background on Vine Copulas} \label{sec:st}

Copulas can be best understood through Sklar's theorem \citep{sklar1959}.
Let $\bm X = (X_1, \dots, X_d) \in \R^d$ be a random vector with joint distribution $F$ and marginal distributions $F_{1}, \dots, F_{d}$.
Then there exists a function $C\colon [0, 1]^d \to [0, 1]$, called \emph{copula}, such that
\begin{align*}
	F(x_1, \dots, x_d) = C(F_1(x_1), \dots, F_d(x_d)), \quad \bm x \in \R^d.
\end{align*}
A copula is itself a distribution function with uniform margins and captures the dependence between the individual random variables.
Now consider the bivariate case for simplicity.
If $X_1$ and $X_2$ are absolutely continuous, we have the following density version of the theorem:
\begin{align}
	f_{1, 2}(x_1, x_2) = f_1(x_1) f_2(x_2)  c_{1, 2}(F_{1} (x_1), F_2(x_2)) , \label{eq:cop-dens}
\end{align}
where $c, f_{1}$,$f_{2}$ are the Lebesgue-densities corresponding to $C$, $F_{1}$, $F_{2}$, respectively.
Further, the copula $C$ is unique and equals the distribution function of the vector $(U_1, U_2) = (F_1(X_1), F_2(X_2))$.

Vine copulas are particularly flexible models for multi-dimensional copulas. They build a multivariate dependence structure using only bivariate copulas as building blocks. 
To ensure the model produces valid and computationally efficient distributions, the bivariate copulas are organized in a graphical structure, called \emph{vine}.
A regular vine (R-vine) is a series of trees $\mathcal{V} = (T_1, \ldots, T_{d-1})$, where each tree is an acyclic graph formed by a set of nodes and edges $(N_t, E_t)$ \citep{bedford2002}.
$\mathcal{V}$ is composed such that the edges in tree $t$ become the nodes of tree $t+1$.
In \citet{bedford2001}, these graph representations are proposed as multivariate probability density functions of factorized conditionally dependent variables.
Rvine copulas are such multivariate conditional dependence structures disassembled into bivariate building blocks where each edge links to a bivariate copula \cite[e.g.]{bedford2001, czado2019, czado2022}.
The resulting decomposed acyclic structure is flexible in higher dimensions and provides a considerable degree of interpretability.
% Refer to \cref{fig:vine5} for an illustrative example.
% In \citet[Lemma 2.4]{kiriliouk2023}, a telescoping is proposed for the factorization of a multivariate  probability density function $f$ of dimension $d > 2$, such that
\citet{bedford2001} have shown that
\begin{equation}\label{eq:vcd}
    f_{1, \dots, d}(\bm x) = \prod_{k = 1}^d f_k(x_k) \cdot \prod_{t = 1}^{d- 1} \prod_{e \in E_t} c_{a_e, b_e; D_e}(x_{a_e | D_e}, x_{b_e | D_e}; \bm x_{D_e}),
\end{equation}
where $x_{i | D_e} = F_{i | D_e}(x_i \mid \bm x_{D_e})$ represents the conditional marginal distribution, and $C_{i, j; D_e}$ denotes the conditional copula of $(X_{i}, X_j)$ given $\bm X_{D_e} = \bm x_{D_e}$. 
Let the multivariate distribution $F$ through the use of vine copulas be described by
\begin{equation}\label{eq:tri}
    F := (\{F_j\}_{j \in J}, \mathcal{V}, \mathbf{C}(\mathcal{V})),
\end{equation}
where $\{F_j\}_{j \in J}$ represents the set of marginal distributions, $\mathcal{V}$ the graph structure of the Rvine, and $\mathbf{C}(\mathcal{V})$ the set of bivariate copulas over all edges within the trees of $\mathcal{V}$.

\cref{eq:vcd} emphasizes that a conditional version of Sklar's theorem is useful for vine copula models \citep[e.g.,]{patton2006}. 
Therefore let $i, j \in \{1, \dots, d\}$ indicate a tuple of variables connected by edge $e$. 
Then assume $D_e = \{1, \dots, d\} \setminus \{i, j\}$, and write $F_{i, j|D_e}$ for the conditional distribution of $(X_i, X_j)$ given $\bm X_{D_e}$ with conditional marginals $F_{i|D_e}, F_{j|D_e}$. 
For every $\bm x_{D_e}$ there is a copula $C_{i,j;D_e}( \cdot ; \bm x_{D_e})$ such that
\begin{align} \label{eq:cond-sklar}
	F_{i,j|D_e}(x_i, x_j | \bm x_{D_e}) = C_{i, j;D_e}(F_{i | D_e} (x_i \mid \bm x_{D_e}), F_j(x_j \mid \bm x_{D_e}); \bm x_{D_e}).
\end{align}
In vine copula models, it is common to employ the \emph{simplifying assumption}  $C_{i, j;D_e}(\cdot ; \bm x_{D_e}) =  C_{i, j;D_e}(\cdot)$ for all $\bm x_{D_e} \in \R^{|D_e|}$. 
Under this assumption, a vine copula model consists of the marginal distributions and one bivariate copula per edge (opposed to one for each value of $\bm x_{D_e}$). 
A downside of the simplifying assumption in vine copulas is that it can overlook genuine, non-constant conditional dependencies among variables. 
However, it allows for fitting each bivariate copula separately, which greatly simplifies both the computational effort and the complexity of the estimation procedure. 
\citet{czado2019} provides evidence that simplifying and non-simplified assumption leads to similar estimates with proofs for specific distributions.
For further details, we refer to \citet{haff2010}, \citet{stoeber2013a}, and \citet{joe2014}.
In our work, the exact validity of the simplifying assumption is of minor importance. Primary interest is in the quality of bias-corrected climate model outputs, which we evaluate extensively \cref{sec:res}, in line with the recommendations of 
\citet{nagler2024b}.

\subsection{Zero Inflated Variables and Other Discrete-continuous Mixtures}\label{sec:zi}

Consider a random variable $X \in \R$ with stochastic representation
\begin{align*}
	X = Z \wt X,
\end{align*}
where $Z \in \mathrm{Bernoulli}(p)$  and $\wt X \in \R$ is another random variable independent of $Z$. 
We shall first deal with the case where $\wt X$ is a positive, absolutely continuous variable.
If the inflation parameter $p > 0$, we call $X$ \emph{zero-inflated}.
The variable $X$ has distribution function
\begin{align*}
	F_X(x) = p\ind(x \ge 0) + (1 - p) F_{\wt X}(x),
\end{align*}
and density (with respect to the sum of the Dirac measure at $0$ and the Lebsgue measure on $\R_{>0}$)
\begin{align*}
	f_X(x) = p\ind(x = 0) + (1 - p)\ind(x > 0)  f_{\wt X}(x).
\end{align*}

Later, we need a generalization of zero-inflated variables. Consider $X \in \R$ with stochastic representation
\begin{align*}
	X = \wt X^{(K)},
\end{align*}
where $\wt X^{(0)} \in \R$ is absolutely continuous, $\wt X^{(k)}, k \ge 1$ are constant variables attaining fixed values $x^{(1)}, x^{(2)}, \dots$, and $K$ is an integer-valued random variable. We call such $X$ a \emph{discrete-continuous mixture} in the following.
Zero-inflated variables are a special case in which $x^{(1)} = 0$ and $K \sim \mathrm{Bernoulli}(p)$.
Now the variable $X$ has distribution function
\begin{align*}
	F_X(x) = \sum_{k = 1}^\infty \Pr(K = k)\ind( x^{(k)} \le x) + \Pr(K = 0) F_{\wt X^{(0)}}(x),
\end{align*}
and density (with respect to the sum of the Dirac measures at $x^{(k)}, k \ge 1,$ and the Lebesgue measure on $\R \setminus \{x^{(k)}, k \ge 1\}$)
\begin{align*}
	f_X(x) = \sum_{k = 1}^\infty \Pr(K = k) \ind( x^{(k)} = x) + \Pr(K = 0)\ind(\nexists \,k \ge 1 \colon  x^{(k)} = x)  f_{\wt X}(x).
\end{align*}

\subsection{A New Vine Decomposition for Mixture Variables} \label{sec:icop}

Recall the density decomposition from \eqref{eq:cop-dens} and that the copula is unique if all variables are continuous.
If a variable is a mixture, things get more complicated.
The copula $C$ is unique only on 
\begin{align*}
	\mathrm{range}(F_1 \times F_2) = \{(u_1, u_2) \in [0, 1]^2\colon \exists (x_1, x_2) \mbox{ with } u_1 = F_1(x_1), u_2 = F_1(x_2)\}.
\end{align*}
This is a minor issue from an application perspective since it suffices to find \emph{any} copula that complies with the data.
Further, a factorization of the density similar to \eqref{eq:cop-dens} can be derived.

\begin{proposition}  \label{prop:mixture-joint}
	For two mixture variables $X_1, X_2$ with copula $C$, define the discrete support $\Xcal_j = \{x \in \R\colon \Pr(X_j = x) > 0\}$, where $F^{-}(x) = \lim_{\epsilon \searrow 0} F(x - \epsilon)$ such that $P(X = x) = F(x) - F^-(x)$ holds. Denote  $C^{(j)}(u_1, u_2) = \partial C(u_1, u_2) / \partial u_j$.
	It holds
	\begin{align*}
		f_{1, 2}(x_1, x_2) & = f_1(x_1) f_2(x_2)  \mathfrak c(x_1, x_2),
	\end{align*}
	with \emph{generalized copula density}
	\begin{align*}
		 & \mathfrak c(x_1, x_2) :=
		\begin{cases}
			\frac{C(F_1(x_1), F_2(x_2)) - C(F_1(x_1), F_2^-(x_2)) - C(F_1^-(x_1), F_2(x_2)) + C(F_1^-(x_1), F_2^-(x_2))}{f_1(x_1)f_2(x_2)}, & x_1 \in \mathcal X_1, x_2 \in \mathcal X_2 \\
			\frac{C^{(2)}(F_1(x_1), F_2(x_2)) - C^{(2)}(F_1^-(x_1), F_2(x_2))}{f_1(x_1)},                                                   & x_1 \in \Xcal_1, x_2 \notin \Xcal_2        \\
			\frac{C^{(1)}(F_1(x_1), F_2(x_2)) - C^{(1)}(F_1(x_1), F_2^-(x_2))}{f_2(x_2)},                                                   & x_1 \notin \Xcal_1, x_2 \in \Xcal_2        \\
			c(F_1(x_1), F_2(x_2)),                                                                                                          & x_1 \notin \Xcal_1, x_2 \notin \Xcal_2,
		\end{cases}
	\end{align*}
	and conditional distribution functions
	\begin{align*}
		F_{1 | 2}(x_1 | x_2) & = \mathfrak h_{1 | 2}(x_1, x_2) := \begin{cases}
			                                                          \frac{C(F_1(x_1), F_2(x_2)) - C(F_1(x_1), F_2^-(x_2))}{f_2(x_2)}, & x_2 \in \Xcal_2     \\
			                                                          C^{(2)}(F_1(x_1), F_2(x_2)),                                      & x_2 \notin \Xcal_2,
		                                                          \end{cases}  \\
		F_{2 | 1}(x_2 | x_1) & = \mathfrak h_{2 | 1}(x_1, x_2) :=  \begin{cases}
			                                                           \frac{C(F_1(x_1), F_2(x_2)) - C(F_1^-(x_1), F_2(x_2))}{f_1(x_1)}, & x_1 \in \Xcal_1     \\
			                                                           C^{(1)}(F_1(x_1), F_2(x_2)),                                      & x_1 \notin \Xcal_1,
		                                                           \end{cases}
	\end{align*}
\end{proposition}

The idea behind the formulas is intuitive. When variables are continuous, we derive the density from the distribution function through differencing. When they are discrete, we take differences. For a mixture, we proceed accordingly depending on whether $x_j$ is a singularity of $X_j$. The proof then amounts to the same computations as in \citet[Section 2.1]{stoeber2013b}.

Note that \cref{prop:mixture-joint} can also be applied in the conditional version of Sklar's theorem with $C =  C_{i, j;D_e}(\cdot ; \bm x_{D_e})$, $F_1 = F_{i |D_e}(\cdot \mid \bm x_{D_e})$, $F_2 = F_{j | D_e}(\cdot \mid \bm x_{D_e})$. This leads to the following generalized vine decomposition result; its proof can be found in \cref{app:proof}.

\begin{proposition} \label{prop:vines}
	Suppose all variables in $\bm X$ are discrete-continuous mixtures.
	\begin{enumerate}[(i)]
		\item The joint density factorizes as
		      \begin{align*}
			      f_{1, \dots, d}(\bm x) = \prod_{k = 1}^d f_k(x_k) \times \prod_{t = 1}^{d- 1} \prod_{e \in E_t} \mathfrak c_{a_e, b_e; D_e}(x_{a_e | D_e}, x_{b_e | D_e}; \bm x_{D_e}),
		      \end{align*}
		      where $x_{i | D_e} = F_{i | D_e}(x_i \mid \bm x_{D_e})$ and $C_{i, j; D_e}$ is the conditional copula of $(X_{i}, X_j)$ given $\bm X_{D_e} = \bm x_{D_e}$.
		\item The conditional distribution functions $F_{j | D}$ can be computed recursively via
		      \begin{align*}
			      F_{j | {D_e}}(x_j \mid \bm x_{D_e}) = \mathfrak h_{j \mid r ; D_e}(x_{j | D_e \setminus{r}}, x_{r \mid D_e \setminus{r}}; \bm x_{D_e\setminus{r}}),
		      \end{align*}
		      where $\mathfrak  h_{j \mid r; D_e}$ is defined as in \cref{prop:mixture-joint} with $C = C_{j, r; D_e}(\cdot; \bm x_{D_e})$, $F_1 = F_{j \mid D_e \setminus r}$, $F_2 = F_{r \mid D_e \setminus r}$.
	\end{enumerate}
\end{proposition}

The density decomposition resembles the formula from the continuous case \eqref{eq:vcd}. This makes it straightforward to adapt the usual computational algorithms \citep[see, e.g.,][]{czado2019} by appropriately substituting the generalized copula densities $\mathfrak c$ and conditionals $\mathfrak h$.

\section{Vine Copula Bias Correction (VBC)} \label{sec:vbc}

\begin{figure}[t]
    \centering
    \includegraphics[width = 0.9\textwidth]{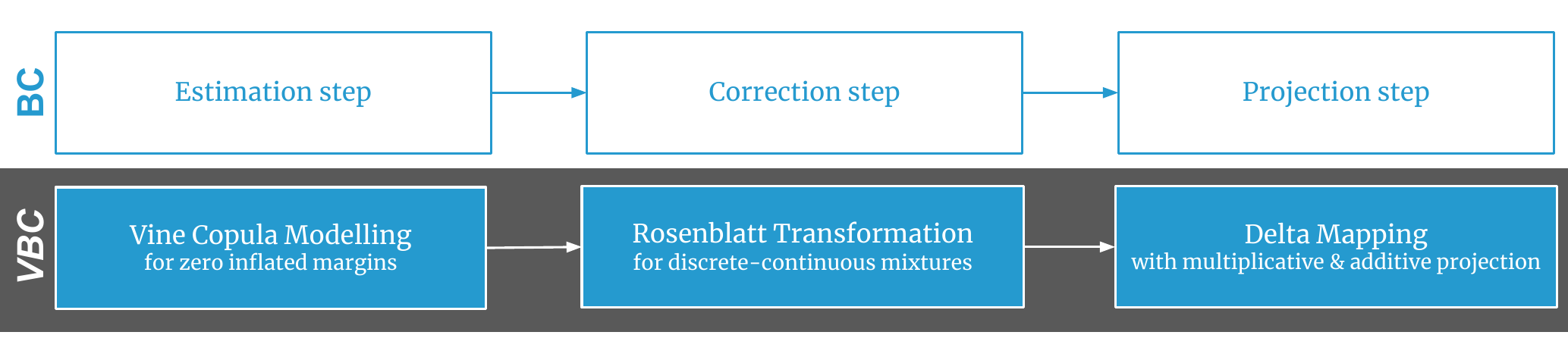}%\rule{438pt}{74pt}
    \caption{The three typical steps of bias corrections (in the upper row) and our proposed solution in \textit{VBC} (in the lower row).}
    \label{fig:bcstep}
\end{figure}

Recall from \cref{sec:bc} that VBC consists of three steps: estimating, correcting, and projecting the climate data.
To estimate the multivariate distribution function of the climate data, VBC employs the proposed concept of zero-inflated regular vine copulas.
VBC corrects the climate model by a specific adaption of the Rosenblatt transformation that can handle non-uniformity in the pseudo observations.
The corrected data are then projected by a delta mapping, that is sensitive for zero-inflated and heavy-tailed marginals.
The procedure is depicted in \cref{fig:bcstep}.
Details for the individual steps are given below.

\subsection{Estimation Step}\label{sec:vce}

The multivariate distribution function of the model and reference is estimated by our former definition of regular vine copulas with zero-inflated margins in \cref{prop:vines}.
The distributions can be written by the three components

\begin{equation*}
\begin{aligned}
    &\mathcal{D}_{rc} \sim & F_{rc} &:= (\{F_{rc,j}\}_{j \in J}, \mathcal V_{rc}, \mathfrak{C}_{rc}(\mathcal V_{rc})),  \\
    &\mathcal{D}_{mp} \sim & F_{mp} &:= (\{F_{mp,j}\}_{j \in J}, \mathcal V_{mp}, \mathfrak{C}_{mp}(\mathcal V_{mp})),   
\end{aligned}
\end{equation*}
from \cref{eq:tri} where $ \mathfrak{C}(\mathcal V)$ decodes the set of bivariate copulas with zero-inflated margins over all edges within the trees of $\mathcal{V}$.

While parametric density estimations are subject to distributional assumptions, kernel density estimation can flexibly and smoothly model various distributional shapes. 
Therefore, the univariate marginals $\{F_{j}\}_{j \in J}$ are estimated using the kernel density estimator implemented in the \textit{kde1d} package \citep{nagler2024} which allows to model for zero-inflated margins.

Interval-scaled variables with continuous densities, such as temperature or dewpoint temperature measured in °C, are estimated by local polynomial kernel density estimation (KDE) \citep{sheather1991}.
The single bounded continuous variable surface wind speed in m/s is estimated using logarithmic transformations \citep{geenens2018}.
For the zero-inflated variables precipitation per m² and shortwave radiation in W/m², we follow the theoretical framework outlined in \cref{sec:zi}. 
Discrete segments in the random variable with mixture distribution are evaluated at the limits of the support $\hat{F}_{X}^-(x^{(k)})$ and $\hat{F}_{X}(x^{(k)})$.
The continuous part is estimated similarly to a bounded variable, with the inflation intervals indicating the respective boundaries.

\begin{figure}[tp]
    \centering
 \begin{tikzpicture}
 \footnotesize
  % Set the radius of the circle nodes
  \def\myradius{0.5cm} % Adjust the size as needed
    
  \begin{scope}[every node/.style={circle,thick,draw,minimum size=\myradius}]
    \node (d) at (0,0.5) {$\texttt{d}$};
    \node (t) at (2,0.5) {$\texttt{t}$};
    \node (p) at (4,0.5) {$\texttt{p}$};
    \node (r) at (6,0) {$\texttt{r}$};
    \node (w) at (6,1) {$\texttt{w}$};
  \end{scope}

  \begin{scope}[every edge/.style={draw=black,thick}]
    \path (d) edge node[above] {$\texttt{dt}$} (t);
    \path (t) edge node[above] {$\texttt{tp}$} (p);
    \path (p) edge node[above] {$\texttt{pr}$} (r);
    \path (p) edge node[above] {$\texttt{pw}$} (w);
  \end{scope}

  \begin{scope}[every node/.style={circle,thick,draw,minimum size=\myradius}]
    \node (dt) at (1,-1) {$\texttt{dt}$};
    \node (tp) at (3,-1) {$\texttt{tp}$};
    mma 2.4], a telescoping is proposed for the factorization of a multivariate
\node (pr) at (5,-0.5) {$\texttt{pr}$};
    \node (pw) at (5,-1.5) {$\texttt{pw}$};
  \end{scope}

  \begin{scope}[every edge/.style={draw=black,thick}]
    \path (dt) edge node[above] {$\texttt{dp} ; \texttt{t}$} (tp);
    \path (tp) edge node[above] {$\texttt{tr} ; \texttt{p}$} (pr);
    \path (tp) edge node[above] {$\texttt{tw} ; \texttt{p}$} (pw);
  \end{scope}
\end{tikzpicture}
    \caption{The figure shows the vine graph structure for a valid factorization. The first line represents graph $T_1$, where the nodes represent the univariate marginal distributions and the edges represent the unconditioned first-order pair copulas. The second line represents graph $T_2$. Note that the edges of $T_1$ represent the set of nodes in $T_2$.}
    \label{fig:vine5}
\end{figure}
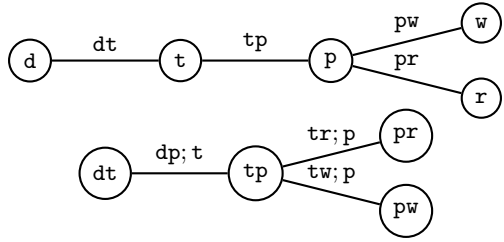

The estimation of vine $\mathcal V$ and the associated pair copulas $ \mathfrak{C}(\mathcal V)$ is implemented using \textit{vinecopulib} \citep{nagler2017}. 
To model the vine $\hat{\mathcal{V}}$, a sequential estimation of its trees $T_1$ to $T_{d-1}$ is necessary.
A tree $T_t$ should represent the (conditional) dependence structure between the climate variables.
\cref{fig:vine5} visualizes an exemplary dependence structure for the first two trees.
The structure can either be determined by expert knowledge or by correlation.
Once determined, the dependence structure identifies the matching pair copulas.
The pair copulas $\mathfrak c_{\texttt{dt}}, \mathfrak c_{\texttt{tp}}, \mathfrak c_{\texttt{pw}}$, and $\mathfrak c_{\texttt{pr}}$ determine the edges and, hence vine graph structure, of the first tree. 
Within the tree, all pair copulas except $\mathfrak c_{\texttt{dt}}$ use the definition of discrete support from \cref{sec:icop}. 

In the estimation process for the vine graph structure, the selection of copula models and the associated parameters are considered as \textit{hyperparameters} of the model.
The number of possible tree structures in a vine is $d! \times 2^{\frac{(d-2)(d-3)}{2}-1}$ \citep{morales2011}.
This results in 480 possible vine structures per estimated multivariate distribution for the five climate variables of interest in \cref{tab:cv}.
For the presented application in \cref{sec:data}, the algorithm provided in \citet{dissmann2013} facilitates the decision on the vine copula structure for the 5-dimensional modelling problem.
Thus, the graph structure is determined tree by tree.
At each level, the highest mutual Kendall's $\tau$ determines the maximal spanning tree.
The pair-copulas are modelled by the nonparametric \textit{local likelihood transformation estimator} (TLL; \citep{nagler2014, geenens2017, geenens2018, loader2006}) only. 
Consequently, there is no need to compare the likelihood of several alternative models using model selection criteria.

\subsection{Correction Step}\label{sec:rt}

The \emph{Rosenblatt transform} \citep{rosenblatt1952} and its inverse are established and essential tools in environmental vine copula model applications \cite[e.g.]{montes2015, sun2021} where it is used for sampling, evaluation or projection purposes.
In Bias Correction, it can be used for the correction step \citep{dekens2017}.
Therefore, the Rosenblatt transform and inverse are applied sequentially to the model data to project them from their native model domain to a targeted reference distribution.

The transformation method turns an absolutely continuous random vector $\bm X = (X_1, \dots, X_d)$ with distribution $F$ into another vector $\bm V = (V_1, \dots, V_d) = R(\bm X)$ containing independent $\mathrm{Uniform}[0, 1]$ variables.
To deal with discreteness, \citet{brockwell2007} proposed a modification of the transform by adding artificial randomness. Let $W_1, \dots, W_d$ be independent  $\mathrm{Uniform}[0, 1]$ random variables that are also independent from $\bm X$, and define
\begin{align*}
	V_{j} =  W_jF_{j | j-1, \dots, 1}(X_j | X_{j - 1}, \dots, X_{1}) +  (1 - W_j) F_{j | j-1, \dots, 1}^-(X_j | X_{j - 1}, \dots, X_{1}),
\end{align*}
where $F^{-}(x | \cdots) = \lim_{\epsilon \searrow 0} F(x - \epsilon| \cdots)$ is the left limit of the conditional distribution.
Theorem 1 of \citet{brockwell2007} shows that $\bm V \sim \mathrm{Uniform}[0, 1]^d$.
The inverse Rosenblatt transform $\bm X = R^{-1}(\bm V)$ turns independent uniform variables $\bm V$ into a vector $\bm X$ with distribution $F$. 
Note that the inverse Rosenblatt transform needs no mechanism to handle discreteness since its input $V$ is strictly continuous.
It is given by,
\begin{align*}
	X_{j} = F_{j | j-1, \dots, 1}^{-1}(V_j | V_{j - 1}, \dots, V_{1}), \quad j = 1, \dots, d.
\end{align*}

The Rosenblatt and inverse Rosenblatt transforms for discrete-continuous mixtures are used to transfer the bias-prone model data from their distributional domain to the more realistic reference domain.
In the first step, the Rosenblatt transform projects an instance of the model data $\bm x_{mp}$ from its native probability distribution $F_{mp}$ onto an independent standard uniform domain $\mathbf{v}_{mp}$:

\begin{equation*}
    R(\bm x_{mp} \mid \hat{F}_{mp}) := \mathbf{v}_{mp},
\end{equation*}
 where $\mathbf{v}_{mp} \sim  \mathrm{Uniform}[0,1]^d$ if $ \hat{F}_{mp}$ is a sufficiently good estimate of the true distribution of the data generating process underlying $\mathcal{D}_{mp}$. 
Subsequently, the uniform climate model sample is projected on the climate reference domain:

\begin{equation*}
    R^{(-1)}(\mathbf{v}_{mp} \mid \hat{F}_{rc}) := \hat{\bm x}_{mc},
\end{equation*}
 where we define  $\hat{\mathcal{D}}_{mc} = \{\hat{\bm x}_{mc}^{(i)}\}_{i = 1}^n$ to be the bias corrected climate data set.
 The Rosenblatt transform scales quadratically in observations and variables.
Note that the data are now bias-corrected as $\hat{\mathcal{D}}_{mc} \sim \hat{F}_{rc}$ holds.
We further claim that the internal course of the weather within the model is preserved if $R(\bm x_{mp} \mid \hat{F}_{mp}) \approx R(\hat{\bm x}_{mc} \mid \hat{F}_{rc})$.

\subsection{Projection Step}\label{sec:dm}

The corrected data $\hat{\mathcal{D}}_{mc}$ represent the course of climate during the projection period, whereas their domain corresponds to the scale of the reference calibration period.
To account for the discrepancy between the correction and projection period, delta mapping restores the differences in the quantiles of the model data.
Let
\begin{align} \label{eq:delta}
    \begin{aligned}
        \Delta_j^* &= \frac{x_{j, mp}}{  \hat{F}_{j, mc}^{-1}(\hat{F}_{j, mp}(x_{j, mp}))} \\
        \Delta_j^+ &= x_{j, mp} - \hat{F}_{j, mc}^{-1}(\hat{F}_{j, mp}(x_{j, mp})) \\
        \hat{x}_{j, mp} &= \begin{cases}
        \hat{x}_{j, mc} \times \Delta_j^*, & \textit{if } \mathbb{P}(X_j < 0) = 0 \textit{ and } \Delta_j^* < 1 \\
        \hat{x}_{j, mc} +  \Delta_j^+, & \textit{else.}
        \end{cases}
    \end{aligned}
\end{align}
The delta mapping in \cref{eq:delta} is used in VBC.
Variables bounded at zero are mapped by a multiplicative difference if $\Delta_j^* < 1$, to avoid negative projections and cope with possible zero inflation.
In any other case, an additive difference maps variables to the projection period.

\textit{Example:} Consider a hypothetical but realistic setup for radiation, wherein the 10\% quantile in the projection period is $x_{r, mp} = F^{-1}_{r, mp}(0.1) = 200$.
The projection is a lot lower $F^{-1}_{r, mc}(0.1) = 10$ and the corrected reference is at $\hat{x}_{j, mc} = 100$
\begin{itemize}
    \item In the multiplicative mapping $\Delta_r^* = 200/10 = 20$ and the correction would suggest $ \hat{x}_{j, mp} = \hat{x}_{j, mc} \times \Delta_j^* =  100 * 20 = 2000$. 
    \item In the additive mapping $\Delta_r^+ = 200-10 = 190$ and the correction would suggest $ \hat{x}_{j, mp} = \hat{x}_{j, mc} \times \Delta_j^+ =  20 + 190= 210$.
\end{itemize}  

The high multiplication factor in the multiplicative mapping suggests an unrealistic value for a lower quantile in radiation. 
Therefore, we suggest using the additive mapping whenever it is possible.
These slight changes in \cref{eq:delta} to delta mapping proposed in UBC \citep{cannon2015} result in a more cautious estimation of discrepancies in lower bounded and heavy-tailed variables.

\section{Application and Results}\label{sec:res}

\subsection{Setup}\label{ssec:su}

In this application, we focus on the bias correction of five climate variables derived from CRCM5-LE and outlined in \cref{tab:cv}.
Correcting those five meteorological variables is key to accurately generating hydrological data for the ClimEx-II project (Climate Change and Hydrological Extreme Events 2nd Phase). 
ClimEx-II is a project investigating the dependence between hydrometeorological extremes and the occurrence of low flows in Bavaria.
For the hydrometeorological extreme analysis in ClimEx-II, the CRCM5-LE data are bias-corrected and downscaled before being used in a hydrological model.
The second phase extends the holistic modeling approach from ClimEx-I proposed in \citet{willkofer2020} by an updated reference data set (see \cref{sec:data}) and the multivariate bias correction presented in this paper.
The selection of a suitable bias correction method is an important choice for a reliable and realistic hydrological model.
Therefore, we compare VBC, the proposed novel bias correction method for partially zero-inflated margins, against a multivariate correction method (MBCn \citep{cannon2018}) and a univariate correction method (UBC \citep{cannon2015}).
The general procedure of VBC is described in \cref{sec:bc}, the detailed underlying methodology is deduced in the following chapters and the methods estimation and hyperparameters are specified in \cref{sec:vce}.
The default arguments recommended by \citet{cannon2023} are used for the compared methods. Ratio quantities are truncated at $0.05$ and, for the multivariate approach, the maximum amount of iterations is set to $30$. 

All three bias correction methods are calibrated using reference data $\mathcal{D}_{rc}$ from 1991 to 2010 and projected onto the subsequent ten years $\mathcal{D}_{mp}$.
To account for the internal variability of the CRCM5 ensemble, $\mathcal{D}_{mc}$ is constituted by the whole ensemble.
This reflects the general ideas implemented and tested in \citet{vaittinada2021}.
The resulting projection period from 2011 to 2020 is available in both, model data $\mathcal{D}_{mp}$ and reference data $\mathcal{D}_{rp}$. 
Thus, the generalization of the corrected model data can be evaluated on the unseen historical reference from 2011 to 2020 $\mathcal{D}_{rp}$.
For a detailed description of the data sets, see \cref{tab:dat}.

The five climate variables outlined in \cref{fig:hydbav} are selected due to their impact on the Bavarian water system.
Thus, they are modelled and corrected jointly with a temporal resolution of 3 hours and a 12 km spatial resolution. 
To account for climatic diversity in Bavaria, biases are adjusted across three distinct catchments: an alpine, a pre-alpine, and a Franconian catchment (see \cref{fig:hydbav}).
The exemplary catchments comprise six to seven 12 km grid cells, each estimated and adjusted individually.

At a temporal resolution of three hours, the scale, amplitude, and dependencies of climate variables are typically strongly associated with the seasonality and the time of the day.
Radiation, as an example, is heavily zero-inflated at night (see \cref{fig:mVr}) and has a non-linear relationship to the temperature at day.
To adequately model those distinct distributional shapes, the translation method in \citet{mpelasoka2009} is adapted and extended.
Thereby, the climate time series is separated into meaningful temporal chunks that are corrected individually.
The day is divided into daytime (6:00 AM to 6:00 PM) and nighttime (6:00 PM to 6:00 AM), and the year into four seasons: winter (December, January, February), spring (March, April, May), summer (June, July, August), and autumn (September, October, November).
To smooth transitions between adjacent temporal chunks, the data are calibrated on overlapping periods.

\textit{Example:} Assume that we want to correct winter (DJF) during nighttime (6:00 PM to 6:00 AM) with a sample size of 4000 instances. In this case, we will sample 1000 instances from the adjacent month and ours, i.e. November and March, 3PM and 9AM. In the estimation step, the added 5000 instances are used to model the vine. Logically, only the inner 4000 instances are corrected and projected.

In total, we apply one correction to each of the 20 grid cells across all 50 members for all four seasons and both daytime and nighttime.
This means a total of 8,000 corrections per tested bias correction method.

With this technical setup given, the corrected data from each method are evaluated for suitability within the ClimEx-II project.
Two evaluation measures are directly derived from the goals of a successful bias correction in \cref{sec:bc} for a quantitative comparison between the BC methods.
Explicitly, a bias correction should maximise the degree of similarity between the reference and corrected data.
\cref{ssec:ds} examines the similarity of the data by the similarity of their multivariate distribution functions. 
Furthermore, the correction method should preserve the course of weather in the model when correcting, which is discussed in \ref{ssec:pw}.
We also evaluate global metrics on the catchments levels in \cref{app:ltm} to asses if average, spatial and temporal properties are sufficiently represented.

\subsection{Distributional Similarity}\label{ssec:ds}

A successful bias correction should project the climate model data from the distribution of the model data onto the distribution of the reference data during the projection period, i.e., $\hat{\mathcal{D}}_{mp} \sim F_{rp}$.
We evaluate this similarity by estimating the second Wasserstein distance $W^2$ \citep{villani2008} between the corrected data $\hat{\mathcal{D}}_{mp} = \{\hat{\bm x}^{(i)}_{mp}\}_{i = 1}^{n_{mp}}$ and the reference data in the projection period $\mathcal{D}_{rp} =\{\bm x^{(j)}_{rp}\}_{j = 1}^{n_{rp}}$ (see \cref{tab:dat}):

\begin{equation}\label{eq:wd}
    W^2(\hat{\mathcal{D}}_{mp}, \mathcal{D}_{rp}) = \left( \min_{\gamma} \sum_{i=1}^{n_{mp}} \sum_{j=1}^{n_{rp}} \gamma_{ij} d(\hat{\bm x}^{(i)}_{mp}, \bm x^{(j)}_{rp})^2 \right)^{1/2},
\end{equation}
where $d$ is the Euclidean distance function and $\gamma$ denotes the transport plan.
If the multivariate data exhibit different scales, the domains of both the model and reference data are standardized by the location and scale of the reference data.
To account for the improvement of a bias correction, the Wasserstein distance between reference and correction from \cref{eq:wd}, is compared to the Wasserstein distance between reference and respective model data
\begin{equation}\label{eq:iwd}
     IW^2(\hat{\mathcal{D}}_{mp} \mid \mathcal{D}_{mp}, \mathcal{D}_{rp}) = W^2(\mathcal{D}_{mp}, \mathcal{D}_{rp}) - W^2(\hat{\mathcal{D}}_{mp}, \mathcal{D}_{rp}).
\end{equation}

This approach quantifies the difference in the effort to transfer the reference distribution to the model distribution and its bias-adjusted equivalent.
A positive $IW^2$ indicates an improvement in distributional similarity after the correction, and a negative $IW^2$ indicates a deterioration.
\\
\begin{figure}[tp]
    \centering
    \includegraphics[width = 0.9\textwidth]{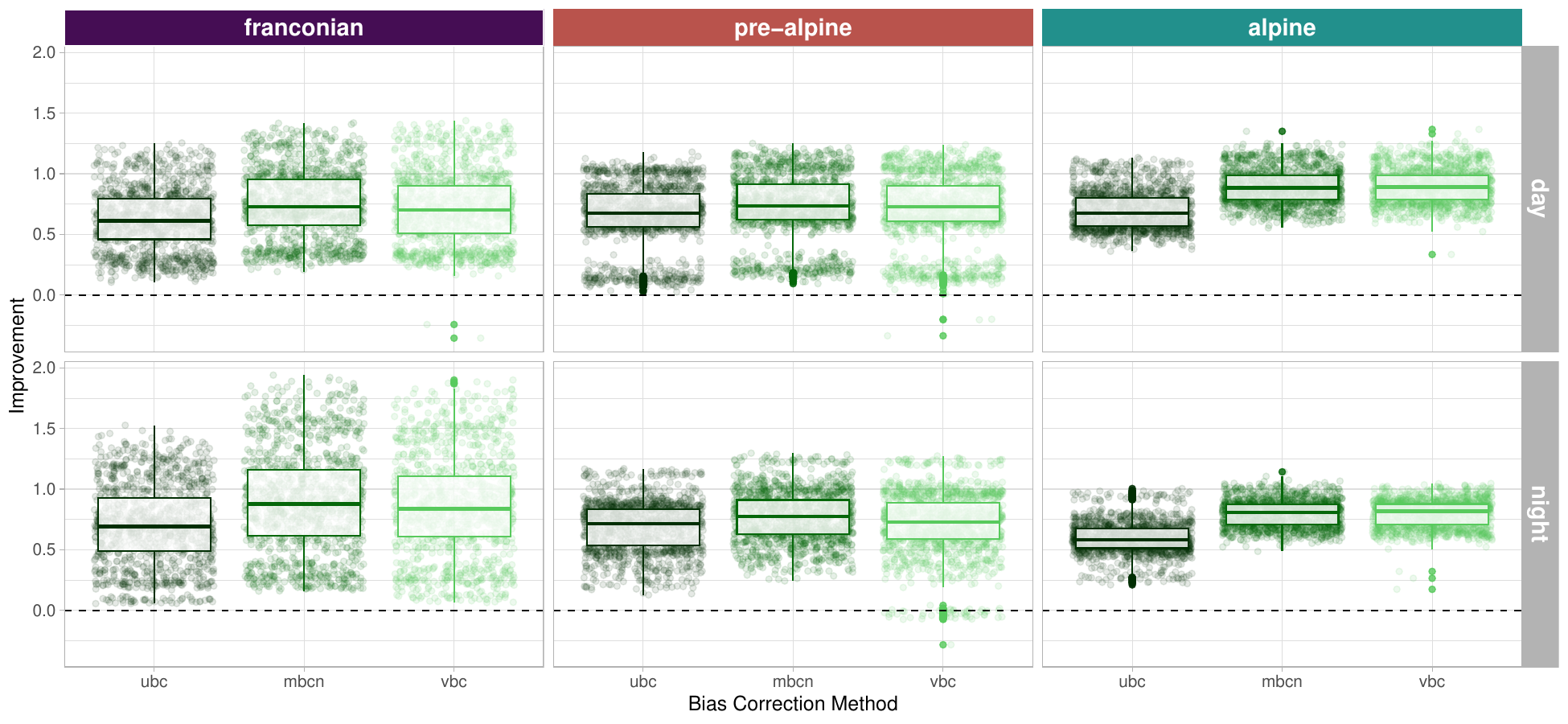}%\rule{438pt}{74pt}
    \caption{Improvement in 2nd Wasserstein distance between the reference data and model data after applying three different bias correction methods spanning 2011 to 2020.
The improvement of each bias correction method is categorised according to the three catchments of \cref{fig:hydbav}: Franconian (purple), pre-alpine (red) and alpine (cyan). 
In the rows day and night time are differentiated.
Individual points are jittered along the x-axis and indicate the improvement for a singular specific correction.
The boxplot indicates the respective distributional quantiles among the points.
Each correction is characterized by an ensemble member and a spatiotemporal subset (see \cref{ssec:su}).
The horizontal line at zero separates corrections that enhanced similarity (above the line) and those that reduced it (below the line).}
    \label{fig:sim_day}
\end{figure}

\cref{fig:sim_day} illustrates the improvement in the 2nd Wasserstein distance $IW^2$ from \cref{eq:iwd} for each adjusted data set.
The evaluation is faceted by the catchment and the day and night.
In all three catchments, in any of the three methods, the bias correction increases the distributional similarity to the reference data in at least 89\% of the corrected data sets (see \cref{tab:imp2time}).
In general, multivariate correction methods exhibit higher correction rates than the univariate method.
Among the three methods and in all the catchments, VBC exhibits the highest improvement rates after bias correction during the day.
In the alpine catchment during daytime, VBC increases the similarity of more than 50\% of the corrections by $IW^2 > 1$.
During nighttime, VBC and MBCn perform similarly in all three catchments.
In the Franconian catchment, VBC improves 75\% of by at least 0.39, MBCn by 0.36 and UBC by 0.3.

\begin{figure}[tp]
    \centering
    \includegraphics[width = 0.9\textwidth]{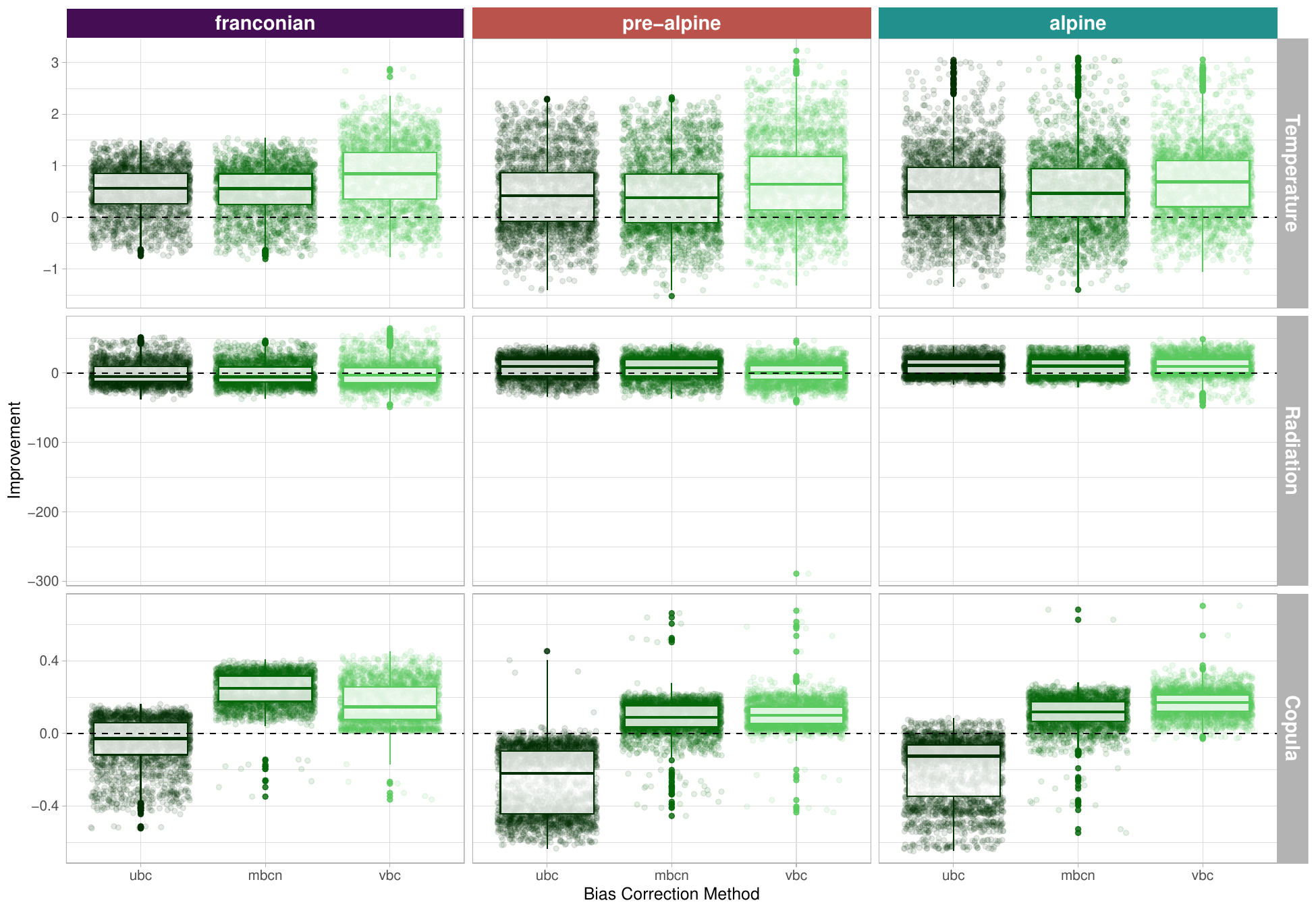}%\rule{438pt}{74pt}
    \caption{Improvement of margin and relationship measured by distributional similarity between the reference data and model data before and after applying three different bias correction methods, spanning the years 2011 to 2020.
Each bias correction method's enhancement is categorised regionally by the three catchments: Franconian (purple), pre-alpine (red), and alpine (cyan).
The two top rows indicate improvements in two analysed margins: temperature distributions (top row) and radiation (middle row).
In the bottom row, the improvement in the distributional similarity between the copulas explains the improvement in the relationship, excluding marginal improvements.
For each row, a jittered point relates to a single correction, characterised by an ensemble member and a spatiotemporal subset (see \cref{ssec:su}).}
    \label{fig:sim_margin}
\end{figure}

\begin{figure}[tp]
    \centering
    \includegraphics[width = 0.85\textwidth]{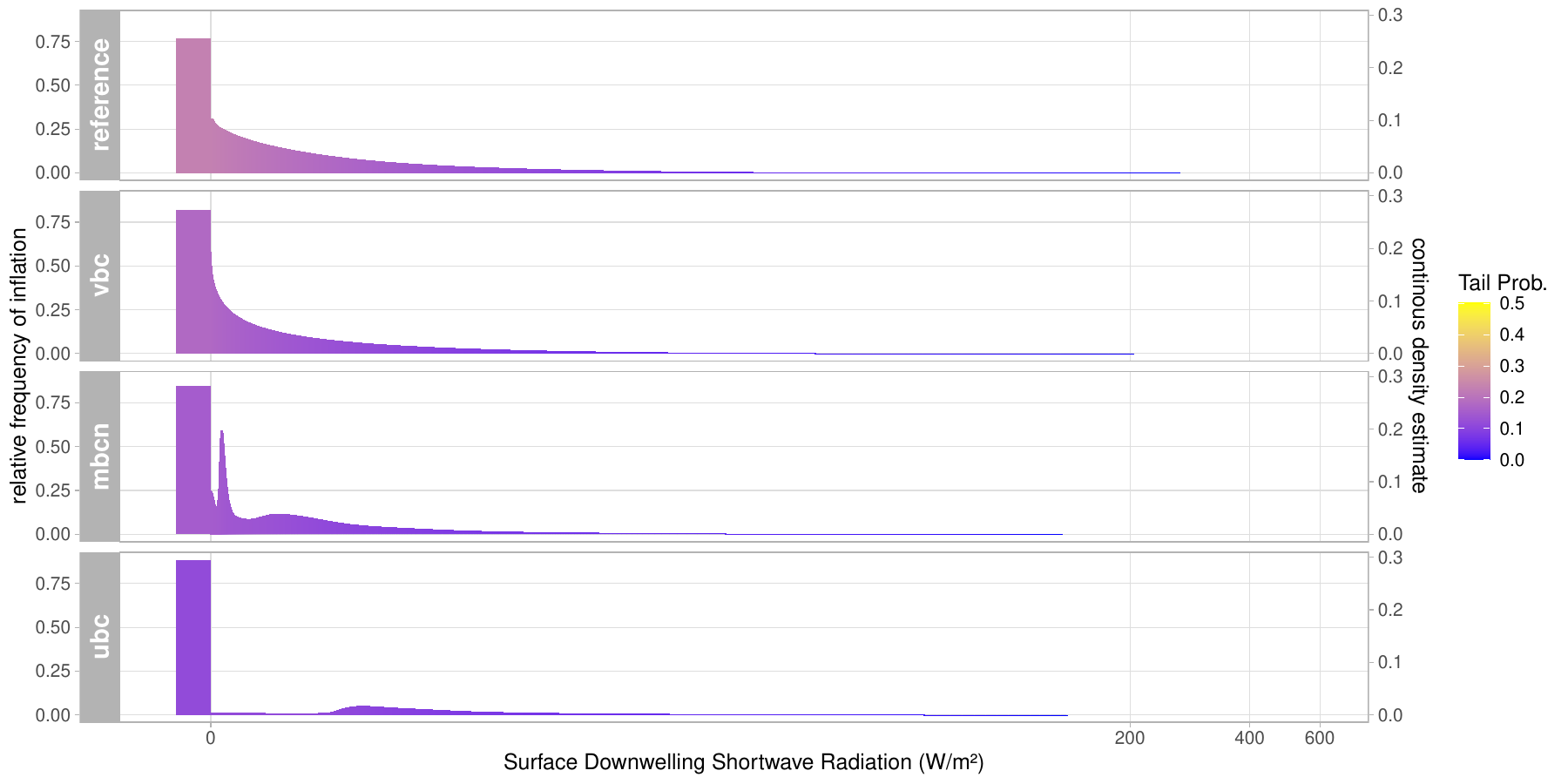}%\rule{438pt}{74pt}
    \caption{Estimated density of the radiation for a grid cell in Iller-Kepmten in the Allgäuer High Alps (47°20'14.64N, 10°12'53.244E) in winter during nighttime between 2011 and 2020.
             In the rows, the densities are faceted by the reference data and the marginal correction results.
             The bar indicates the relative inflation of the discrete, and the density curve represents the continuous domain of the climate variable.
             For comparison, the colour gradient encodes the tail probabilities of radiation with yellow encoding the median and blue encoding the tails.
             The domain is scaled by $ln(x+1)$ for better interpretability of the heavy tail.}
    \label{fig:pix_rsds}
\end{figure}

For further intuition, \cref{fig:sim_margin} depicts the improvement in the 2nd Wasserstein distance for the univariate margins of the climate variables temperature (top row) and radiation (mid row) and the empirical copula (bottom row).
The results for the other margins, precipitation, wind, and dewpoint, can be found in \cref{tab:imp2marg}.
Note that, different from \cref{fig:sim_day}, no standardisation is needed since the copula extends over the uniform hypercube $[0,1]^d$ and the margins are univariate.
For the margin of temperature, VBC has the highest median rates of improvement in the Franconian and alpine catchment. 
VBC shows equal correction quality with UBC of $IW^2 \approx 0.45$ in the pre-alpine catchment.
MBCn corrects the univariate margin worse than the other two methods in all 3 catchments.
For the radiation margin, all three corrections have a median of around zero, with no clear trend of improvement visible.
This might be due to the fact that the distribution of radiation seems to change between the calibration and the projection period.
A more detailed description of the correction of radiation can be found in \cref{fig:pix_rsds}.
For the empirical copula, the multivariate correction methods MBCn and VBC improve the similarity to the reference data in 88\%--100\% and 93\%--100\%, respectively (see \cref{tab:imp2marg}).
The univariate correction UBC, on the other hand, harms the distributional similarity in 94\% of the corrections in the Ziller catchment and in 99\% of the Iller catchments.
VBC and MBCn exhibit comparable median correction strength at the pre-alpine and alpine catchments.
At the Franconian catchment, MBCn has a slightly higher median correction strength than VBC.
Overall, UBC exhibits a deterioration in the similarity in the relations between the variables.
VBC has the highest chance of improving the multivariate copula of the model data (94\%). 

For all three BC methods, the correction of the radiation margins is the most challenging.
Therefore, radiation is exemplary compared in one pre-alpine grid cell in the Allgäuer High Alps, during winter nights.
In this grid cell, the density curves in \cref{fig:mVr} indicate strong discrepancies in radiation between the reference and the uncorrected model data. 
\cref{fig:pix_rsds} compares the marginal densities of radiation between the corrected and the reference data.
At the discrete, inflated part of the radiation, VBC, MBCn, UBC, and the reference have comparable quantiles of 75\% - 80\% zero inflation.
In the continuous domain of radiation, VBC underestimates the heavy tail, while MBCn and UBC exaggerate the density mass in higher value ranges.
The curvature of the margin corrected by VBC is very similar to the curvature of the reference.
In the case of MBCn, the curvature has a steep mode at ca. 10 W/m² and a rather flat but long tail.
UBC and MBCn exhibit a rather complex, multi-modal curvature over their heavy tail.
Both corrections fail to reflect the slope and scale of the reference data.

As shown in Appendix \ref{app:res}, all three bias-correction methods effectively adjust for central tendency, as well as temporal and spatial structures, thereby validating the correction procedure.
Given the application in \cref{sec:data}, VBC and MBCn are the most powerful bias correction methods among the three evaluated.
VBC has the highest chance of improving the multivariate relationships between the variables.
For all three methods, radiation is hard to correct for.
However, the distribution corrected by VBC in \cref{fig:pix_rsds} looks the most realistic and smooth. 
Moreover, the proposed method shows the highest certainty among the compared methods to improve the similarity of temperature post-correction.
The comparatively good results in the depiction of the margins in \cref{fig:sim_margin} and the multivariate distribution in \cref{fig:sim_day} suggest that VBC finds a good compromise in correcting the multivariate distribution function and margins. 
Specifically, if relationships between the variables are important to be corrected, VBC is the best method among the three compared.

\subsection{Preservation of Weather}\label{ssec:pw}

When correcting for bias in the climate model data, univariate quantile mapping implicitly preserves the rank of the climate variable from the climate model \citep{cannon2015}.
The correction step transfers the model to the reference data by
\begin{equation}\label{eq:pmci}
    \hat{x}_{mc, j}^{(t)} := F_{rc, j}^{-1}(F_{mp, j}(x_{mp,j}^{(t)})) \quad \text{such that} \quad  F_{rc, j}(\hat{x}_{mc, j}^{(t)}) = F_{mp, j}(x_{mp,j}^{(t)}),
\end{equation}
where $t$ refers to the time step that is corrected.
If, for example, an extreme precipitation occurs at time $t$ in the uncorrected model, the event is still as extreme after correction.
By preserving the climate model's inherent temporal rank structure, the course of the climate variable within the model data is maintained after correction.
In multivariate correction, we do not aim to maintain the course of one climate variable, but the \textit{course of weather}, characterised by the multivariate structure of multiple climate variables.
Multivariate bias corrections adjust the composition of ranks within a multivariate vector.
For example, if an extreme event like a snowstorm occurs at time $t$ in the uncorrected model, the course of this snowstorm should be maintained after correction.
This course involves an interplay between multiple variables, especially dewpoint temperature, air temperature, and precipitation.
In terms of its scale, however, the snowstorm should ideally look like a snowstorm in the reference data set.
The consequence of multivariate correction is that the rank structure is no longer preserved, i.e., $F_{rc, j}(\hat{x}_{mc, j}^{(t)}) \neq F_{mp, j}(x_{mp, j}^{(t)})$.
Further discrepancies between calibration and projection might obfuscate relationships in the rank structure.

To evaluate if a multivariate correction preserves the course of weather obtained from the model data, we introduce the \textit{Model Correction Inconsistency (MCI)}.
Similar to preserving the course of singular climate variables in the univariate correction, we propose to compare the temporal structure of the non-exceedance probabilities to maintain the course of weather in the multivariate case.
Non-exceedance probabilities are the probabilities derived from the multivariate distribution function, i.e., the probability that a random weather event is less or equally extreme (see, e.g., \citep{zscheischler2018, du2015}).
The non-exceedance probability of an event $t$ can be measured by
\begin{equation*}
    F_{\mathcal{D}}(\bm x^{(t)}) = \mathbb{P}(X_1 \leq x^{(t)}_1, \ldots, X_d \leq x^{(t)}_d).
\end{equation*} 
$F_{\mathcal{D}}$ denotes the multivariate distribution function of the data set $\mathcal{D}$, and $\bm x^{(t)} = (x^{(t)}_1, \ldots, x^{(t)}_d)$ is a realization of $\mathcal{D}$ at time $t$.
This is the logical \textit{AND} event, which measures the probability that none of the climate variables exceeds the observation at $t$.
Other non-exceedance probabilities like the Kendall or Survival Kendall distribution function might also be appropriate alternatives here \citep{salvadori2007, zscheischler2018}.
We consider the course of weather to be preserved if a climate event exhibits equal non-exceedance probabilities before and after correction.
Thus, we define the Model-Correction-Inconsistency as the absolute difference of the non-exceedance probability of a weather event $t$ between the model data and its bias-corrected sibling,
\begin{equation}\label{eq:mcits}
    MCI^{(t)} = \mid F_{\mathcal{D}_{mp}}(\bm x^{(t)}_{mp}) - F_{\hat{\mathcal{D}}_{mp}}(\hat{ \bm x}^{(t)}_{mp}) \mid,
\end{equation}
where $\hat{\mathcal{D}}_{mp} = \{\hat{\bm x}^{(t)}\}_{i = 1}^n$ denotes the corrected data set $\mathcal{D}_{mp}$.
Given that $MCI^{(t)}  \in [0, 1]$, $0$ indicates the desirable non-invasive correction that preserves the course of weather at $t$.
$1$ indicates a perfectly invasive correction that alters the course of weather to the opposite extremity.
For a global estimation of the inconsistency, the average MCI is used:
\begin{equation}\label{eq:gmci}
  MCI = \frac{1}{n} \sum_{t = 1}^n MCI^{(t)}
\end{equation}

The MCI is derived from the rank-preserving property of UBC in \cref{eq:pmci}.
Consequently, UBC can be interpreted as a benchmark.
If a multivariate correction results in an MCI close to the MCI of UBC, this can be understood as a non-invasive correction.
An MCI of exactly zero is only realistic under specific circumstances that are independent of the correction method, e.g. if the calibration and the projection period are equal.
We propose that a threshold of $MCI < 0.05$, indicates no clear invasive signal, i.e. the correction method preserves the non-exceedance probabilities of the model.

\begin{figure}[tp]
    \centering
    \includegraphics[width = 0.9\textwidth]{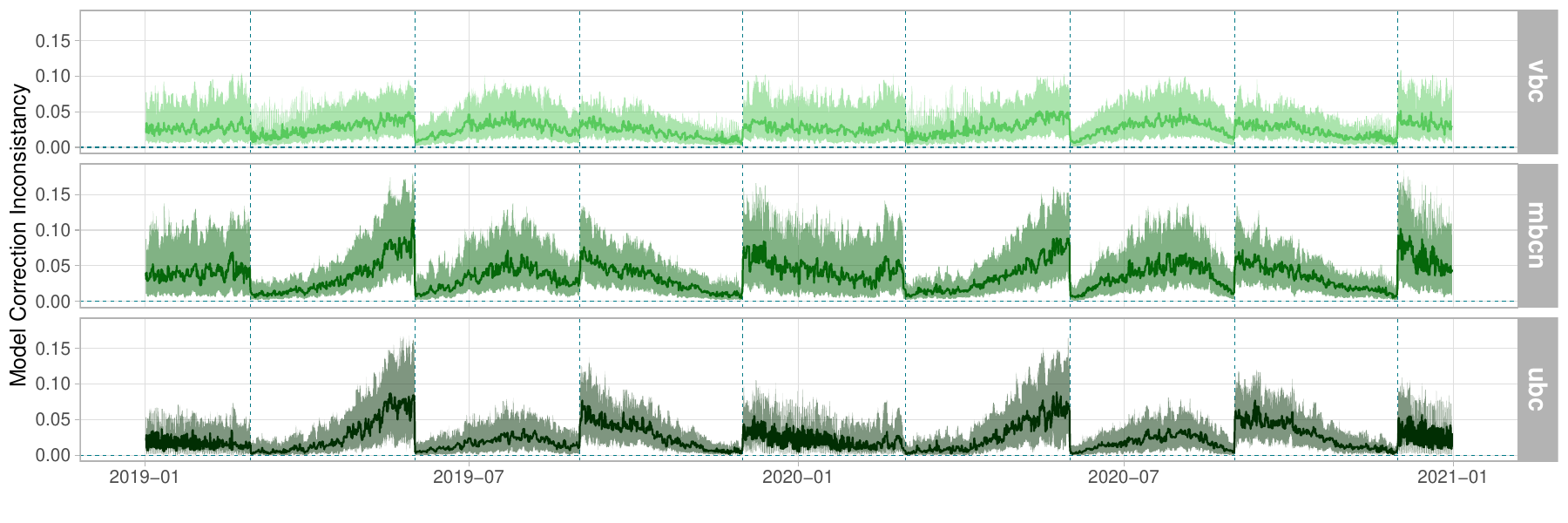}
   \caption{Model-Correction-Inconsistency time series $MCI^{ (t)}$ at the exemplary pre-alpine grid cell in the Allgäuer High Alps (47°20'14.64N, 10°12'53.244E) in the most recent years 2019 and 2020.
    Each row depicts the $MCI^{ (t)}$ for one of the three bias corrections.
    In each panel, the solid line depicts the median time series over all 50 ensemble members and the shaded background indicates the 25\% and 75\% quantiles.
    The blue horizontal line indicates the desirable non-invasive correction at $MCI^{(t)} = 0$.
    The vertical lines mark the four seasons- winter, spring, summer and autumn.}
  \label{fig:mci_time}
\end{figure}

\cref{fig:mci_time} summarises the $MCI^{ (t)}$ from \cref{eq:mcits} over all 50 ensemble members of CRCM5 from 2019 and 2020 for each of the three compared bias correction methods for the exemplary grid cell in the Allgäuer High Alps.
Among the three methods, the time series of VBC and UBC is the closest to zero with an average $MCI^{(t)} $ of 0.02, indicating that the least invasive BC method.
For the years 2019 and 2020, MBCn in the middle row has an average $MCI^{(t)} $ of 0.04. 
Given the underlying climate model data in \cref{fig:mci_time}, the vine copula correction VBC in the top panel has a better ability to preserve the course of weather than the compared multivariate correction of MBCn.
MBCn has the strongest average inconsistency and the highest variance in the correction.
Further, all three methods exhibit seasonal patterns in the $MCI^{(t)}$, visible by the shifts in $MCI^{(t)}$ between the seasons and the curvature of the slopes within the seasons.
The shifts in $MCI^{(t)}$ between the seasons are an artefact of the translation method, examined in \cref{ssec:su}.
The smaller the shift, the smoother the implementation of the translation method.
Visibly, the highest shifts occur for MBCn with an average change of almost 0.063 in $MCI^{(t)}$.
UBC reacts less sensitively to the translation method with average shifts of ca. 0.055, and VBC handles the translations the smoothest with an $MCI^{(t)}$ of 0.028. 
The curvature of the $MCI^{(t)}$ within the seasons indicates cyclic inconsistencies between the uncorrected and corrected course of weather.   
At the beginning of spring, the time series for VBC, MBCn and UBC are close to zero.
Towards the end of spring, all three methods tend to interfere more with the original course of weather.
MBCn elevates to a median inconsistency of over 0.01 during May.
In the middle of summer in July, MBCn seems to interfere more strongly with weather events.
The maximum recorded $MCI^{ (t)}$ in the depicted period is 0.17, scored by MBCn at the beginning of September 2019.
The corresponding maxima of VBC and UBC are 0.1 and 0.16.
For the example visualised in \cref{fig:mci_time}, MBCn is the most invasive correction method.
The preservation rate of VBC is comparable to that of the univariate method.

\begin{figure}[tp]
    \centering
    \includegraphics[width = 0.99\textwidth]{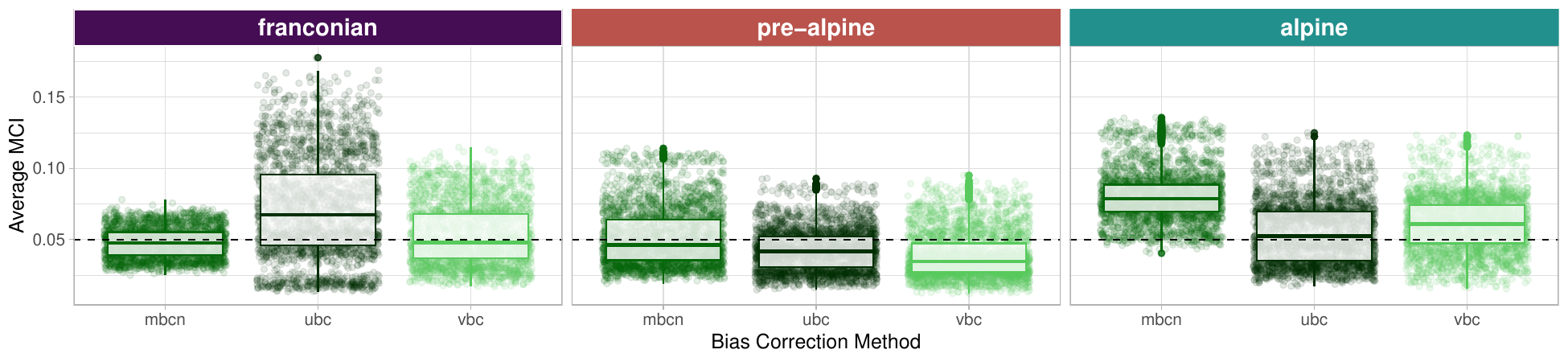}%\rule{438pt}{74pt}
    \caption{Depiction of the average MCI for the three compared BC methods during the projection period from 2011 to 2020.
             The figure is faceted by the Franconian, the pre-alpine and the Alpine catchment (from left to right).
             A jittered point relates to evaluating the MCI for a single correction, characterised by an ensemble member and a spatiotemporal subset (see \cref{ssec:su}).
             Points below the threshold of 0.05 are categorised as having no invasive signal. 
             The box plots summarise the distribution of the jittered points.
             }
    \label{fig:mci_catch}
\end{figure}

\cref{fig:mci_catch} visualizes the average MCI from \cref{eq:gmci} by the BC method and the respective Bavarian catchment.
In the Franconian catchment, VBC and MBCn are the least invasive corrections, having the lowest median inconsistency of 0.05 (see \cref{tab:mci2catch}).
The MCI of UBC is characterised by a huge difference in this catchment, with an inter-quartile range between 0.05 and 0.1. 
At the pre-alpine (Iller) catchment, VBC is the least invasive BC method with a median MCI of 0.03.
At the alpine (Ziller) catchments, UBC and VBC are the least invasive methods, with an MCI of 0.05 and 0.06, respectively,  followed by MBCn with an MCI of 0.08.
VBC and UBC are able to correct more than 50\% of the observations without a clear invasive signal, that is, $MCI < 0.05$. 
MBCn, however, corrects only 38\% of the corrections with no invasive signal.
Specifically in the alpine catchment, MBCn cannot preserve the model-specific weather pattern.
For most corrections in VBC and UBC, the variability of the model data can be maintained after correction since the member-specific relative course of the weather is similar before and after correction.

To summarise, UBC and VBC preserve the variability of the model by scoring low in the MCI.
According to \cref{fig:mci_time}, MBCn has a high seasonal variation in the local $MCI^{(t)}$.
Among the two multivariate methods, \cref{fig:mci_catch} indicates that VBC preserves more or at least equal variability in terms of the MCI than MBCn for most corrections in all three catchments.

\section{Discussion}

Physical and hydrological climate variables often exhibit pronounced asymmetric dependence structures \citep{bacigal2011}. 
The VBC framework could be strengthened by broadening its set of candidate copula families and adopting more sophisticated vine architectures. 
In particular, integrating families capable of asymmetry, such as Archimax copulas \citep{charpentier2014}, holds great promise to improve the fit of vine copula models to climate model output.

Since Regional Climate Models do not reproduce spatial dependencies correctly \citep{bardossy2012}, there is a need to check for spatial and temporal dependencies in the corrected data.
Although the presented method, VBC, does not explicitly incorporate spatial dependencies, the results in \cref{app:ltm} indicate that there is a correction factor for the spatial dependencies.
However, some BC techniques incorporate an explicit space and/or time adjustment into the correction procedure \citep[e.g.]{vrac2015, mehrotra2016, nahar2018, mehrotra2021}. 
For VBC, spatio-temporal dependencies could be integrated naturally through the vine structure.
This might be a promising extension in future work.
In the simplest scenario, a fixed vine structure can be applied to all grid cells, bypassing the algorithm proposed by \citet{dissmann2013}. However, more sophisticated approaches are also feasible, where the vine model learns the underlying spatial structure inherent in the data \cite[e.g.]{erhardt2015}.
Although these more advanced solutions offer greater flexibility, they may come with increased computational costs.

Computational costs are typically much higher for multivariate methods than for univariate methods.
Compared to UBC, the costs of VBC are $\approx$ 25--35 times higher, while the costs of MBCn are $\approx$ 20--30 times higher in the large ensemble application \citep{cannon2018}.
However, VBC offers multiple possibilities to reduce the complexity of a larger number of features and observations.
A discussion on the parametrisation of VBC is drawn out in \cref{sec:vce}.
The main hyperparameters of VBC are the set of possible copula families and the vine's dependence structure.
The Dissmann algorithm is used across many application domains of vine copulas to cope with the growing complexity of the vine structure \citep{dissmann2013, allen2017}.
To enhance the computation time, the following steps might be helpful in the proposed order:
\begin{enumerate}
    \item Parallelization: If cores are accessible, the estimation process can be parallelized, since the bivariate copulas in each tree can be optimized separately.
    \item Family set: The family set can be reduced to TLL and the independence copula. The TLL copula is the most flexible among the set of possible families, and optimising TLL only is less complex than evaluating multiple families.
    \item Truncation: If the dimensionality grows too high, the tree structure may be truncated to avoid the computation of less relevant conditional dependencies at the end of the vine structure. Since the Dissmann algorithm selects the most important (conditional) dependencies at each tree structure, the most essential dependencies in terms of correlation should be captured at higher tree structures.  
\end{enumerate}

The delta mapping in the correction step preserves the multivariate dependence structure modelled by the vine copula structure, copula families, and respective parameters $F:= (\{F_j\}_{j \in J}, \mathcal{V}, \mathfrak{C}(\mathcal{V}))$ from the reference data during the calibration period.
This means that the VBC projection implicitly assumes stationarity in the dependence structure.
\citet{vrac2022} show that this assumption of stationarity can be meaningful.
However, for some use cases, the stationarity assumption can be problematic.
Since the delta mapping in VBC is detached from estimation and projection, any other projection step could be plugged in here to be used instead of the proposed method.

\section{Conclusion}\label{sec:dis}

When correcting the bias in large ensemble climate models with high daily or sub-daily resolution, the multivariate distribution of climate variables often exhibits zero inflation and heavy tails.
In this work, we proposed VBC, a bias correction that models the multivariate dependencies of high-resolution climate data using vine copula models.
Vine copulas are known for their flexibility in modelling high-dimensional, nonlinear dependencies and, hence, are an appropriate method to model, quantify and analyse climate compound events \citep{zscheischler2018, zscheischler2020}.
Constructed from the most granular (marginal) level to more complex bivariate relationships, the layers of information in vine copulas result in an interpretable joint distribution \citep{mcneil2015}.
While vine copulas have been used before in the downscaling and bias correction literature \citep{piani2012, sun2021}, our work is the first to address zero inflation in climate variables adequately.
To this end, we generalised the vine decomposition of multivariate distributions to allow for zero-inflated variables.
Using the Rosenblatt transform, such models are then used to correct and project the model simulations to a more realistic reference distribution.

In a real-world application, we compare our new approach to a univariate baseline and a multivariate state-of-the-art method in three central European catchments with distinct climates.
We find that VBC has the highest chance of improving the multivariate dependence in the model data after correction among all three bias correction methods.
Regarding improvement in strength, MBCn and VBC outperform the univariate correction method.
Exemplary results indicate a potential deficit in the methodology of VBC for correcting heavy tails in specific marginal distributions.
If the correct representation of extremes in univariate margins is of greater relevance, the kernel density estimator in higher quantiles (e.g. above 0.95 or 0.99) could be replaced by suitable extreme value distributions.
However, VBC reproduces smoother, more realistic density representations of zero-inflated margins than MBCn and UBC.

To ensure that the specific patterns of the used model data are still maintained after correction, we introduce the \textit{Model Correction Inconsistency (MCI)}.
The MCI is a multivariate metric that compares the data before and after correction to quantify whether the relative course of weather in the model is preserved after correction.
Compared to MBCn, VBC has fewer seasonal divergences and is less invasive overall.
This indicates that VBC is less sensitive to the translation method used in the application.
Among the three methods, only UBC and VBC correct more than 50\% of the model data with no invasive signal.

In summary, the newly proposed VBC method is the most suitable correction for correcting multivariate sub-daily climate simulations.
The method is the only one to correct the multivariate climate model data to a realistic level while maintaining the course of weather that the simulation model originally provided, at least in our application.
We believe that the potential of our contributions goes beyond mere bias correction and could be exploited in other research fields like economic applications \citep{lambert1992, yip2005, shi2018}, epidemiology \citep{preisser2012, bohning1999} or other climate research.
Either the capabilities of modelling zero inflation for data sets with high temporal resolution and/or the potential to correct multivariate distributions to target distributions while maintaining the original course of the data could be used or adopted in fields of climate research, like downscaling, hydrology research, and compound assessment.
The packages of vinecopulib \citep{nagler2017} and VBC \citep{funk2024b} provide a powerful code base for implementations and extensions of zero-inflated vine copula modelling and vine-based correction of multivariate distributions, respectively.

\newpage

\appendix

\section{Proof of \texorpdfstring{\cref{prop:vines}}{Proposition \ref*{prop:vines}}} \label{app:proof}

\begin{enumerate}[(i)]
	\item For some set $S \subseteq \{1, \dots, d\}$, denote $f_S(\bm x) = f_S(\bm_S)$ as the density of $(\bm X_s)$.
	      By the vine telescoping product formula \citep[Lemma 2.4]{kiriliouk2023}, it holds
	      \begin{align*}
		      f_{1, \dots, d}(\bm x) = \prod_{k = 1}^d f_k(x_k) \times \prod_{t = 1}^{d- 1} \prod_{e \in E_t} \frac{f_{\{a_e, b_e\} \cup D_e}(\bm x)}{f_{\{a_e\} \cup D_e}(\bm x) f_{\{b_e\} \cup D_e}(\bm x)}.
	      \end{align*}
	      Further, using
	      \begin{align*}
		      f_{S | S'}(\bm x) = \frac{f_{S}(\bm x)}{f_{S'}(\bm x)}, \quad S' \subsetneq S,
	      \end{align*}
	      we obtain
	      \begin{align*}
		      \frac{f_{\{a_e, b_e\} \cup D_e}(\bm x)}{f_{\{a_e\} \cup D_e}(\bm x) f_{\{b_e\} \cup D_e}(\bm x)} =  \frac{f_{a_e, b_e | D_e}(\bm x)}{f_{\{a_e\} \cup D_e}(\bm x) f_{b_e | D_e}(\bm x)}.
	      \end{align*}
	      Applying \cref{prop:mixture-joint} with $C = C_{a_e, b_e;D_e}(\cdot; \bm x_{D_e})$, $F_1 = F_{a_e | D_e}(\cdot; \bm x_{D_e})$, $F_2 = F_{b_e | D_e}(\cdot; \bm x_{D_e})$, yields
	      \begin{align*}
		      \frac{f_{\{a_e, b_e\} \cup D_e}(\bm x)}{f_{\{a_e\} \cup D_e}(\bm x) f_{\{b_e\} \cup D_e}(\bm x)} = \mathfrak c_{a_e, b_e | D_e}(\bm x),
	      \end{align*}
	      as claimed.

	\item It holds
	      \begin{align*}
		      F_{j|D}(x_j \mid \bm x_D)
		       & = \lim_{\eps \searrow 0} \frac{F_{j, r|D \setminus r}(x_j, x_r \mid \bm x_{D\setminus r}) - F_{j, r|D \setminus r}(x_j, x_r - \eps \mid \bm x_{D\setminus r}) }{F_{r|D \setminus r}(x_r \mid \bm x_{D\setminus r}) - F_{ r|D \setminus r}(x_r - \eps \mid \bm x_{D\setminus r})}.
	      \end{align*}
	      If $x_r \in \Xcal_r$, it holds
	      \begin{align*}
		      \lim_{\eps \searrow 0} F_{r|D \setminus r}(x_r \mid \bm x_{D\setminus r}) - F_{ r|D \setminus r}(x_r - \eps \mid \bm x_{D\setminus r}) = f_{r|D \setminus r}(x_r \mid \bm x_{D\setminus r}).
	      \end{align*}
	      Further, using the conditional version Sklar's theorem \cref{eq:cond-sklar},
	      \begin{align*}
		       & \quad \lim_{\eps \searrow 0} F_{j, r|D \setminus r}(x_j, x_r \mid \bm x_{D\setminus r}) - F_{j, r|D \setminus r}(x_j, x_r - \eps \mid \bm x_{D\setminus r})                                 \\
		       & = C_{j, r|D \setminus r}(x_{j | D}, x_{r|D} \mid \bm x_{D\setminus r}) - C_{j, r|D \setminus r}(x_{j | D}, F_{ r|D \setminus r}^-(x_r\mid \bm x_{D\setminus r}) \mid \bm x_{D\setminus r}).
	      \end{align*}
	      The last two displays together with \cref{prop:mixture-joint} and $C = C_{h, r;D}(\cdot; \bm x_{D})$, $F_1 = F_{j | D}(\cdot; \bm x_{D})$, $F_2 = F_{r | D}(\cdot; \bm x_{D})$ yield
	      \begin{align*}
		      F_{j|D}(x_j \mid \bm x_D) =  \mathfrak h_{j \mid r ; D}(x_{j | D \setminus{r}}, x_{r \mid D \setminus{r}}; \bm x_{D\setminus{r}}).
	      \end{align*}
	      If on the other hand $x_r \notin \Xcal_r$, it holds
	      \begin{align*}
		      \lim_{\eps \searrow 0} \frac{ F_{r|D \setminus r}(x_r \mid \bm x_{D\setminus r}) - F_{ r|D \setminus r}(x_r - \eps \mid \bm x_{D\setminus r})}{\eps} = f_{r|D \setminus r}(x_r \mid \bm x_{D\setminus r}),
	      \end{align*}
	      and
	      \begin{align*}
		       & \quad \lim_{\eps \searrow 0} \frac{F_{j, r|D \setminus r}(x_j, x_r \mid \bm x_{D\setminus r}) - F_{j, r|D \setminus r}(x_j, x_r - \eps \mid \bm x_{D\setminus r}) }{\eps}                                                            \\
		       & = \lim_{\eps \searrow 0} \frac{ C_{j, r|D \setminus r}(x_{j | D}, x_{r|D} \mid \bm x_{D\setminus r}) - C_{j, r|D \setminus r}(x_{j | D}, F_{ r|D \setminus r}(x_r - \eps\mid \bm x_{D\setminus r}) \mid \bm x_{D\setminus r})}{\eps} \\
		       & =  C^{(2)}_{j, r|D \setminus r}(x_{j | D}, x_{r|D} \mid \bm x_{D\setminus r})  f_{ r|D \setminus r}(x_r\mid \bm x_{D\setminus r}),
	      \end{align*}
	      so that
	      \begin{align*}
		      F_{j|D}(x_j \mid \bm x_D)  = C^{(2)}_{j, r|D \setminus r}(x_{j | D}, x_{r|D} \mid \bm x_{D\setminus r}) = \mathfrak h_{j \mid r ; D}(x_{j | D \setminus{r}}, x_{r \mid D \setminus{r}}; \bm x_{D\setminus{r}}),
	      \end{align*}
	      as claimed. \qedsymbol
\end{enumerate}

\section{Aggregated location and trend metrics}\label{app:ltm}

In the following, three global metrics are evaluated to assess the correction performance on the catchment level.
The median values over time and space, the daily lag structure and the spatial autocorrelation. 
Each metric is compared between the desirable reference, a pre-correction metric and the three correction methods.

\begin{figure}[tp]
    \centering
    \includegraphics[width = 0.99\textwidth]{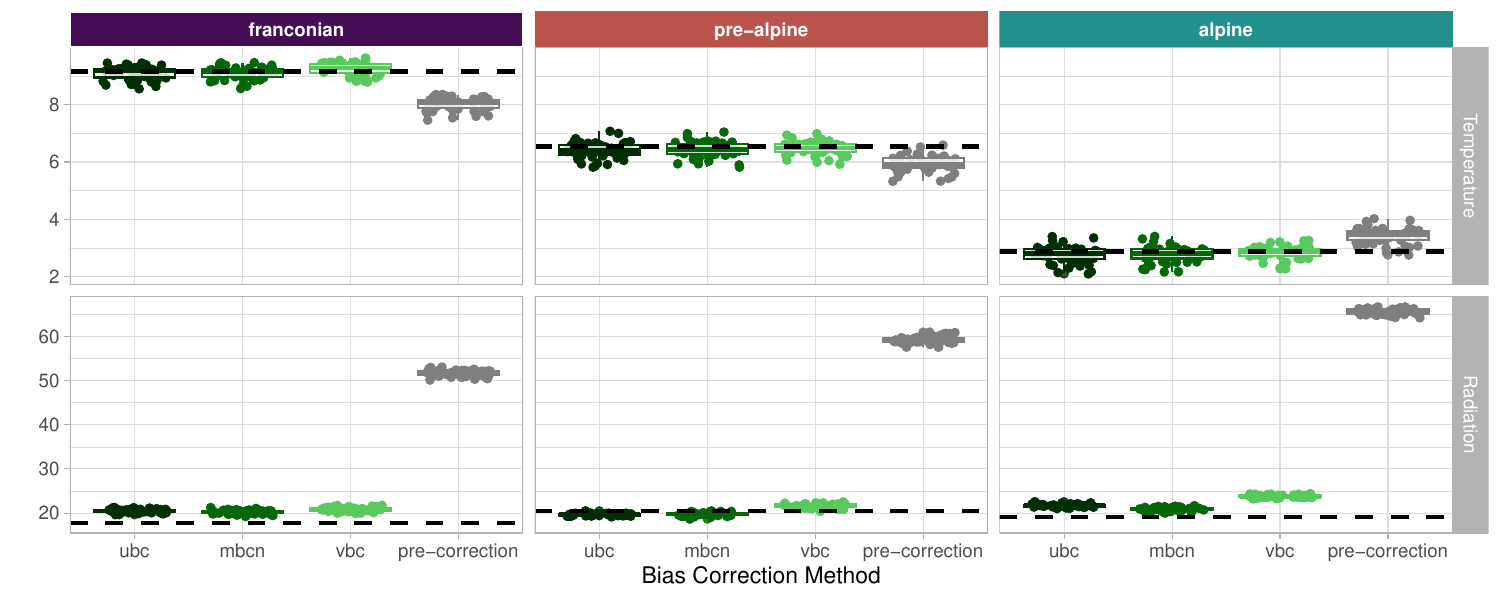}%\rule{438pt}{74pt}
    \caption{Median value for reference and pre-correction model statistics and the three compared BC methods during the projection period from 2011 to 2020.
             The figure is faceted by the Franconian, the pre-alpine and the Alpine catchment (from left to right) and temperature and radiation (in the rows).
             The dotted line depicts the median statistic for the reference.
             Each jittered point depicts a single model member.
             The box indicates the distribution of the members in the CRCM5-LE.
             }
    \label{fig:mean_catch}
\end{figure}

\cref{fig:mean_catch} evaluates the median statistics for Temperature and Radiation in each of the three catchments.
We compare the mean statistic pre-correction (in grey) and the three post-correction results (in green) with the reference mean statistic indicated by the dashed line.
The closer the corrected median is to the reference, the more accurate the corrected ensemble is.
For temperature, the pre-correction CRCM5-LE exhibits a cold bias in the Franconian and pre-alpine catchment and a slight warm bias in the alpine catchment.
In all corrections and for all catchments, the inner 50\% quantile, indicated by the box, covers the reference statistic.
The highest deviations from the median occur in the pre-alpine catchment.
The median correction for the 50 members of VBC differs by a maximum of 0.1, and for MBCn and UBC it is 0.11 and 0.14, respectively (\cref{tab:avg2catch}).
The correction rate in the radiation median is higher than that for temperature. 
In the alpine catchment, the median radiation of the model ensemble is three times higher than that of the reference.
All three BC methods in all three catchments correct the bias to a maximum remaining bias of $\pm 3.1$.
In general, all three methods adjust the bias in temperature and radiation towards the reference median.
We argue that the entire distribution should be assessed for a more specific quality evaluation and comparison of the methods.
Therefore, refer to \cref{ssec:ds}.

\begin{table}[ht]
    \tiny
\centering
\begin{tabular}{rllrrrrr}
  \hline
 climatology & margin & \textbf{pre-correction} & \textbf{UBC} & \textbf{MBCn} & \textbf{VBC} & \textbf{reference} \\ 
  \hline
  & \textbf{temperature} &  8.06 & 9.16 & 9.14 & 9.26 & 9.16 \\ 
  \textcolor{saale}{\textbf{franconian}} & \textbf{radiation} & 51.82 & 20.45 & 20.22 & 20.81 & 17.82 \\ 
  \hline
  & \textbf{temperature} & 5.93 & 6.40 & 6.43 & 6.49 & 6.54 \\ 
  \textcolor{iller}{\textbf{pre-alpine}} & \textbf{radiation} & 59.21 & 19.66 & 19.77 & 21.66 & 20.52 \\ 
  \hline
  & \textbf{temperature} & 3.46 & 2.81 & 2.82 & 2.88 & 2.89 \\ 
  \textcolor{ziller}{\textbf{alpine}} & \textbf{radiation} & 65.47 & 21.69 & 20.87 & 23.80 & 19.06 \\ 
   \hline
\end{tabular}
\caption{Mean statistic by catchment for temperature and Precipitation before and after correction and for reference.
This table incorporates information about the median from \cref{fig:mean_catch}.} 
\label{tab:avg2catch}
\end{table}

\begin{figure}[tp]
    \centering
    \includegraphics[width = 0.99\textwidth]{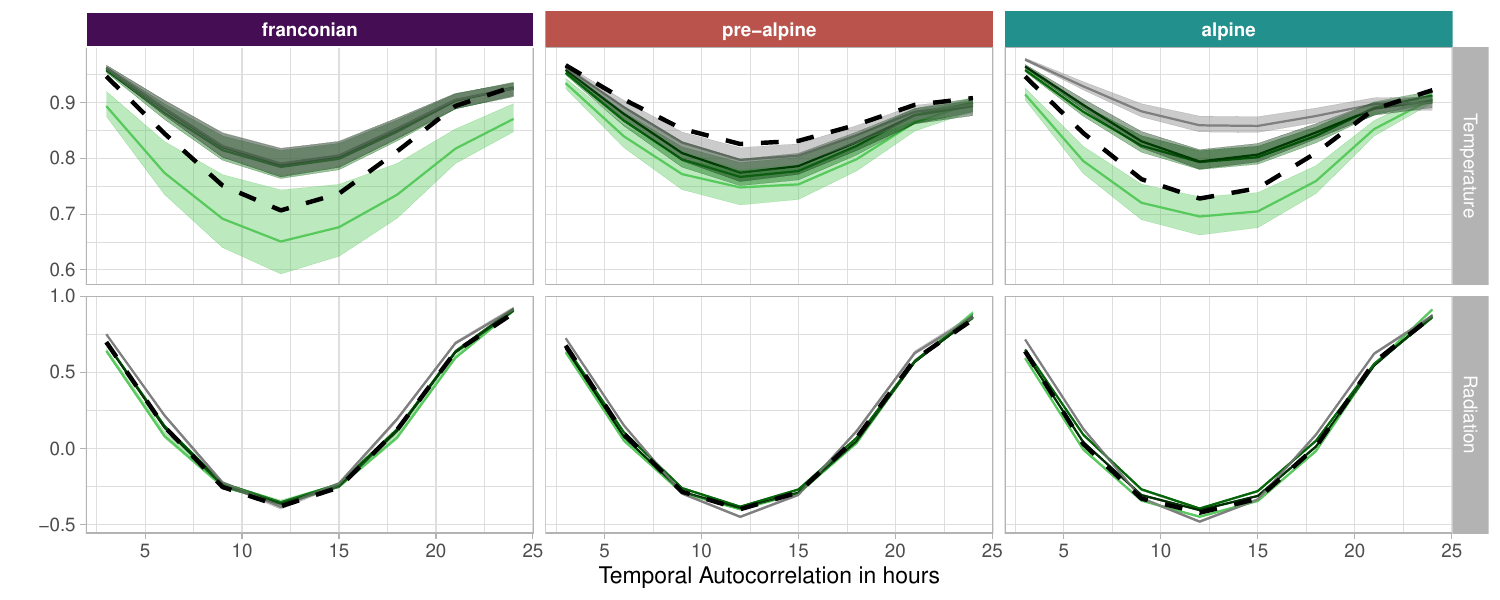}%\rule{438pt}{74pt}
    \caption{Diurnal autocorrelation structure for reference and pre-correction model statistics and the three compared BC methods during the projection period from 2011 to 2020.
             The figure is faceted by the Franconian, the pre-alpine and the Alpine catchment (from left to right) and temperature and radiation (in the rows).
             The plot distinguishes between the mean lag structure of uncorrected model data (grey line) and three bias correction methods: UBC, MBCn, and VBC, shown with dotted lines in a green gradient from dark to light.
             The dotted line depicts the lag structure for the reference.}
    \label{fig:lag_catch}
\end{figure}

\cref{fig:lag_catch} evaluates the diurnal lag structure for temperature and radiation in the three assessed catchments.
The closer the model lag structure is to the black line, the more realistic the representation of diurnal autocorrelation structure by the model.
In the Franconian and Alpine catchments, the autocorrelation structure of the temperature in the model is higher than in the reference.
In the pre-alpine catchment, it is vice versa.
UBC and MBCn do not alter the lag structure of the CRCM55-LE temperature and therefore maintain the lag bias.
This can be seen particularly clearly in the Franconian catchment area, where the lag structure overlaps almost perfectly.
VBC exhibits lower autocorrelation in the three catchments than the CRCM5-LE.
All three correction methods never miss the temperature lag structure of the reference by more than 0.09 (see \cref{tab:lag2catch}).
For radiation, the lag structure of all compared pre- and post-corrections aligns closely with the reference.
Therefore, we argue that, in general, the representation of the lag structure by the corrections is adequate. 

\begin{figure}[tp]
    \centering
    \includegraphics[width = 0.99\textwidth]{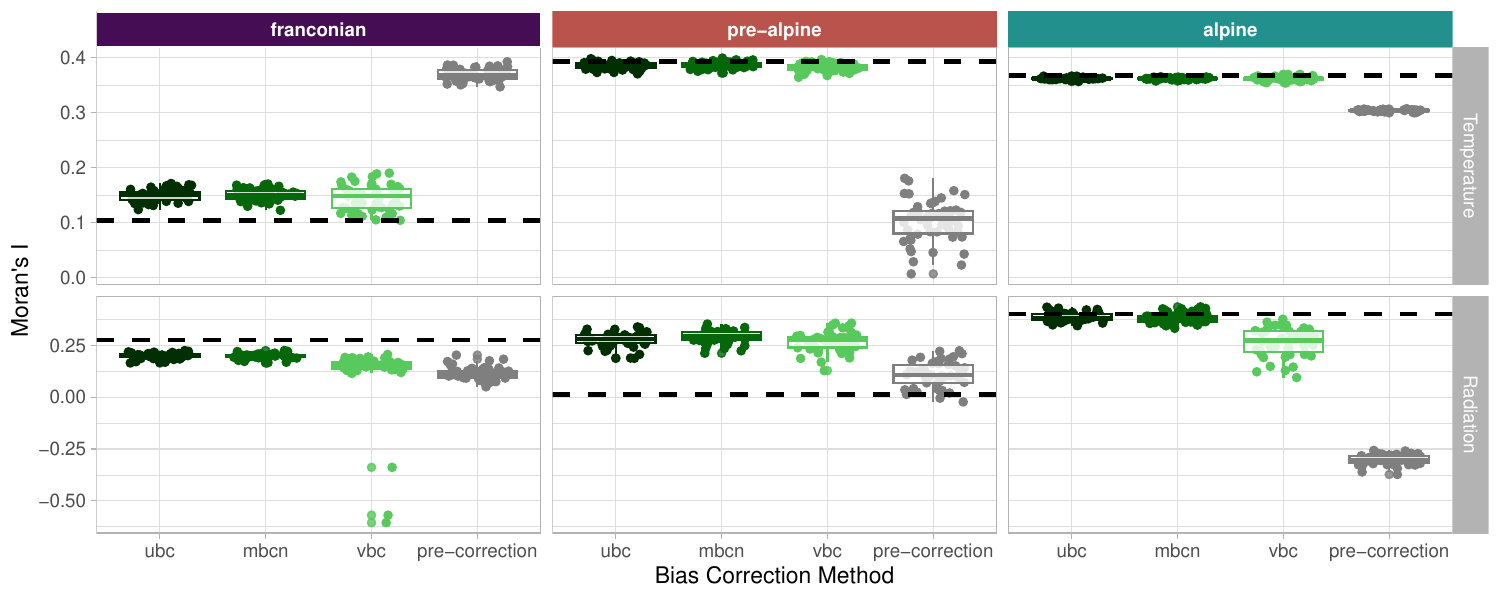}%\rule{438pt}{74pt}
    \caption{Spatial autocorrelation structure for reference and pre-correction model statistics and the three compared BC methods during the projection period from 2011 to 2020.
             The figure is faceted by the Franconian, the pre-alpine and the Alpine catchment (from left to right) and temperature and radiation (in the rows).
             The dotted line depicts the spatial statistic for the reference set.
             A jittered point relates to the average of a single ensemble member.
             The box plots summarise the distribution of the jittered points.}
    \label{fig:moran_catch}
\end{figure}

\cref{fig:moran_catch} evaluates the spatial autocorrelation structure for temperature and radiation in the three assessed catchments.
Moan's I measures the spatial correlation. 
The respective weight matrix accounts for the direct neighbours.
The closer the models' spatial autocorrelation structure over all 50 members, indicated by the boxplots,  to the black line, the more realistic the representation of spatial autocorrelation by the model.
In all three catchments, the bias correction methods enhance the spatial representation of temperature compared to the model data in grey (\cref{tab:avg2moran}).
For radiation, all correction methods enhance the spatial similarity to the reference in the Franconian and the Alpine catchment.
There is a consistent degradation of the spatial representation of radiation over all three correction methods.
This might be due to different climatology in the calibration climate (1991-2010) and the projection climate (2011-2020).
All three corrections implicitly enhance the spatial autocorrelation structure of the model.
For this catchment-based hydrometeorological analysis, we argue that the correction methods sufficiently adjust for the spatial autocorrelation of the data.

\begin{table}[ht]
\tiny
\centering
\begin{tabular}{rllrrrrr}
  \hline
 climatology & margin & \textbf{reference} & \textbf{UBC} & \textbf{MBCn} & \textbf{VBC} & \textbf{pre-correction} \\ 
  \hline
   & \textbf{temperature} & 0.37 & 0.15 & 0.15 & 0.15 & 0.10 \\ 
  \textcolor{saale}{\textbf{franconian}} & \textbf{radiation} & 0.11 & 0.20 & 0.20 & 0.15 & 0.28 \\ 
  \hline
   &  \textbf{temperature} & 0.11 & 0.39 & 0.39 & 0.38 & 0.39 \\ 
  \textcolor{iller}{\textbf{pre-alpine}} & \textbf{radiation} & 0.11 & 0.28 & 0.29 & 0.28 & 0.01 \\ 
  \hline
   &  \textbf{temperature} & 0.30 & 0.36 & 0.36 & 0.36 & 0.37 \\ 
  \textcolor{ziller}{\textbf{alpine}} & \textbf{radiation} & -0.30 & 0.38 & 0.38 & 0.27 & 0.40 \\ 
   \hline
\end{tabular}
\caption{Spatial Autocorrelation by catchment for temperature and Precipitation before and after correction and for reference.
This table incorporates information about the median from \cref{fig:moran_catch}.} 
\label{tab:avg2moran}
\end{table}

The catchment-based analysis evaluates exemplary aggregated metrics on the variables temperature and radiation.
In most cases, the three compared correction methods adjust adequately for median, spatial and temporal autocorrelation.
This confirms the appropriateness of the application setup from \ref{ssec:su}.

\section{Results in Tables}\label{app:res}

The subsequent tables support figures and explanations from \cref{sec:res} with additional information.
\cref{tab:imp2time} and \cref{tab:imp2marg} substantiate the evaluations of distributional similarity before and after bias correction in \cref{ssec:su}.
\cref{tab:mci2catch} substantiates the evaluations in the course of weather by the three correction methods in \cref{ssec:pw}.

\begin{table}[H]
\tiny
\centering
\begin{tabular}{lllrrrr}
  \hline
  \textbf{climatology} & \textbf{diurnal} & \textbf{measure} & \textbf{$W^2$ CRCM5} & \textbf{$IW^2$ UBC} & \textbf{$IW^2$ MBCn} & \textbf{$IW^2$ VBC} \\ 
  \hline
  &  & median & 1.59 & 0.73 & 0.87 & 0.97 \\ 
    &  & 25\% quantile & 1.36 & 0.49 & 0.63 & 0.55 \\ 
    &  & 75\% quantile & 1.87 & 0.95 & 1.14 & 1.04 \\ 
    \textcolor{saale}{\textbf{franconian}} & \textbf{day} & \% of imp. & 0 & 1.00 & 1.00 & 1.00 \\ 
    \cline{2-7}
    & \textbf{night} & median & 1.87 & 0.63 & 0.74 & 0.68 \\ 
    &  & 25\% quantile & 1.64 & 0.30 & 0.36 & 0.39 \\ 
    &  & 75\% quantile & 2.16 & 0.99 & 1.00 & 0.91 \\ 
    &  & \% of imp. & 0 & 0.89 & 0.94 & 0.93 \\
    \hline
    &  & median & 1.62 & 0.83 & 0.89 & 0.88 \\ 
    &  & 25\% quantile & 1.50 & 0.71 & 0.78 & 0.77 \\ 
    &  & 75\% quantile & 1.81 & 0.95 & 1.05 & 1.04 \\ 
    \textcolor{iller}{\textbf{pre-apline}} & \textbf{day} & \% of imp. & 0 & 1.00 & 1.00 & 1.00 \\ 
    \cline{2-7}
    & \textbf{night} & median & 1.86 & 0.68 & 0.79 & 0.72 \\ 
    &  & 25\% quantile & 1.72 & 0.46 & 0.52 & 0.52 \\ 
    &  & 75\% quantile & 2.02 & 0.81 & 0.90 & 0.86 \\ 
    &  & \% of imp. & 0 & 1.00 & 1.00 & 0.98 \\ 
    \hline
    &  & median & 1.87 & 0.82 & 1.00 & 1.02 \\ 
    &  & 25\% quantile & 1.79 & 0.69 & 0.89 & 0.91 \\ 
    &  & 75\% quantile & 1.96 & 0.95 & 1.14 & 1.13 \\ 
    \textcolor{ziller}{\textbf{alpine}} & \textbf{day} & \% of imp. & 0 & 1.00 & 1.00 & 1.00 \\ 
    \cline{2-7}
    & \textbf{night} & median & 1.94 & 0.56 & 0.74 & 0.75 \\ 
    &  & 25\% quantile & 1.79 & 0.44 & 0.61 & 0.60 \\ 
    &  & 75\% quantile & 2.02 & 0.67 & 0.84 & 0.84 \\ 
    &  & \% of imp. & 0 & 1.00 & 1.00 & 1.00 \\ 
   \hline
\end{tabular}
\caption{Distributional information for the improvement of Wasserstein distance in the 8000 corrections, faceted by the climatology of the respective catchment and by day and nighttime.
This table incorporates information from \cref{fig:sim_day}.
In $measure$, the 25\%, 50\% and 75\% quantiles are reported.
Additionally, the percentage of improvements reports the relative amount of improved corrections, i.e., the percentage of corrections where $IW^2 > 0$.
The $IW^2$ columns report the distributional information of the Improvement in 2nd Wasserstein distance for the three correction methods.
$W^2$ displays the distributional information of the 2nd Wasserstein distance between the model and reference as a baseline.} 
\label{tab:imp2time}
\end{table}

\begin{table}[H]
\tiny
\centering
\begin{tabular}{lllrrrr}
  \hline
  \textbf{climatology} & \textbf{margin} & \textbf{measure} & \textbf{$W^2$ CRCM5} & \textbf{$IW^2$ UBC} & \textbf{$IW^2$ MBCn} & \textbf{$IW^2$ VBC} \\  
  \hline
   &    & median & 0.54 & -0.03 & 0.25 & 0.16 \\ 
     & \textbf{copula} & \% of imp. & 0.00 & 0.41 & 1.00 & 1.00 \\ \cline{2-7}
     &  & median & 0.72 & 0.12 & 0.13 & 0.12 \\ 
    & \textbf{dewpoint} & \% of imp. & 0.00 & 0.65 & 0.65 & 0.64 \\ \cline{2-7}
     &  & median & 0.31 & 0.07 & 0.11 & 0.12 \\ 
   \textcolor{saale}{\textbf{franconian}} & \textbf{precipitation} & \% of imp. & 0.00 & 0.80 & 0.94 & 0.92 \\ \cline{2-7}
     &  & median & 26.76 & -5.10 & -5.23 & -6.65 \\ 
     & \textbf{radiation} & \% of imp. &  0.00 & 0.37 & 0.37 & 0.27 \\ \cline{2-7}
     &  & median & 1.37 & 0.87 & 1.11 & 0.95 \\ 
    & \textbf{windspeed} & \% of imp. &  0.00 & 0.99 & 1.00 & 1.00 \\ \cline{2-7}
     &  & median & 1.53 & 0.63 & 0.63 & 0.73 \\ 
     & \textbf{temperature} & \% of imp. & 0.00 & 0.83 & 0.82 & 0.84 \\ 
  \hline
      &  & median & 0.41 & -0.23 & 0.09 & 0.07 \\ 
      & \textbf{copula} & \% of imp. & 0.00 & 0.01 & 0.88 & 0.96 \\ \cline{2-7}
      &  & median & 1.17 & 0.47 & 0.46 & 0.52 \\ 
      & \textbf{dewpoint} & \% of imp. & 0.00 & 0.78 & 0.78 & 0.83 \\ \cline{2-7}
      &  & median &  0.57 & 0.19 & 0.25 & 0.22 \\ 
   \textcolor{iller}{\textbf{pre-alpine}} & \textbf{precipitation} & \% of imp. & 0.00 & 0.80 & 0.87 & 0.80 \\ \cline{2-7}
      &  & median & 33.90 & 9.39 & 7.38 & -4.74 \\ 
      & \textbf{radiation} & \% of imp. & 0.00 & 0.65 & 0.64 & 0.38 \\ \cline{2-7}
      &  & median & 1.70 & 1.21 & 1.31 & 1.35 \\ 
      & \textbf{windspeed} & \% of imp. & 0.00 & 0.97 & 1.00 & 0.99 \\ \cline{2-7}
      &  & median &  1.33 & 0.47 & 0.42 & 0.46 \\ 
      & \textbf{temperature} & \% of imp. & 0.00 & 0.71 & 0.68 & 0.72\\ \cline{2-7}
  \hline
     &  & median & 0.47 & -0.13 & 0.12 & 0.15 \\ 
     & \textbf{copula} & \% of imp. & 0.00 & 0.06 & 0.92 & 0.99  \\ \cline{2-7}
     &  & median & 1.70 & 0.69 & 0.67 & 0.66 \\ 
     & \textbf{dewpoint} & \% of imp. & 0.00 & 0.71 & 0.71 & 0.72 \\ \cline{2-7}
     &  & median & 0.47 & 0.08 & 0.14 & 0.11 \\ 
   \textcolor{ziller}{\textbf{alpine}} & \textbf{precipitation} & \% of imp. & 0.00 & 0.63 & 0.72 & 0.67 \\ \cline{2-7}
     &  & median &  40.27 & 10.53 & 9.66 & 3.52 \\  
     & \textbf{radiation} & \% of imp. &  0.00 & 0.73 & 0.69 & 0.59 \\ \cline{2-7}
     &  & median &  1.89 & 1.24 & 1.33 & 1.41 \\ 
     & \textbf{windspeed} & \% of imp. & 0.00 & 1.00 & 1.00 & 1.00 \\ \cline{2-7}
     &  & median &  1.20 & 0.55 & 0.53 & 0.56 \\ 
     & \textbf{temperature} & \% of imp. & 0.00 & 0.77 & 0.76 & 0.79 \\ 
\end{tabular}
\caption{Distributional information for the improvement of Wasserstein distance in margins and copula, faceted by the climatology of the respective catchment.
This table incorporates information from \cref{fig:sim_margin}.
In $measure$, the  50\% quantile is displayed and the percentage of improvements reports the relative amount of improved corrections.
The $IW^2$ columns report the distributional information of the Improvement in 2nd Wasserstein distance for the three correction methods.
$W^2$ displays the distributional information of the 2nd Wasserstein distance between the model and reference as a baseline.} 
\label{tab:imp2marg}
\end{table}

\begin{table}[H]
\tiny
\centering
\begin{tabular}{llrrr}
  \hline
 climatology & measure & $MCI$ MBCn & $MCI$ UBC & $MCI$ VBC \\ 
  \hline
    & median &  0.0477 & 0.0673 & 0.0479 \\ 
    \textcolor{saale}{\textbf{franconian}} & 25\% quantile & 0.0390 & 0.0459 & 0.0369 \\ 
    & 75\% quantile & 0.0552 & 0.0957 & 0.0682 \\ 
    \hline
    & median &  0.0463 & 0.0418 & 0.034 \\ 
    \textcolor{iller}{\textbf{pre-alpine}} & 25\% quantile & 0.0355 & 0.0306 & 0.0271 \\ 
    & 75\% quantile & 0.0639 & 0.0522 & 0.0473 \\ 
   \hline
    & median & 0.0788 & 0.0525 & 0.0609 \\ 
    \textcolor{ziller}{\textbf{alpine}} & 25\% quantile & 0.0697 & 0.0351 & 0.0473 \\ 
    & 75\% quantile & 0.0886 & 0.0695 & 0.0742 \\ 
   \hline
\end{tabular}
\caption{Distributional information for the Model Correction Inconsistency in the 8000 corrections, faceted by the climatology of the respective catchment.
This table incorporates information from \cref{fig:mci_catch}.
In $measure$, the 25\%, 50\% and 75\% quantiles are reported.
The $MCI$ columns report the distributional information of the preservation of weather for the three correction methods.} 
\label{tab:mci2catch}
\end{table}

\begin{table}[ht]
\centering
\tiny
\begin{tabular}{rllrrrrrr}
  \hline
  climatology & margin & lag & \textbf{pre-correction} &  \textbf{UBC} & \textbf{MBCn} & \textbf{VBC} & \textbf{reference} \\ 
  \hline
  &  & 3.00 & 0.96 & 0.96 & 0.96 & 0.89 & 0.95 \\ 
  &  & 6.00 & 0.89 & 0.89 & 0.88 & 0.77 & 0.84 \\ 
  &  & 9.00 & 0.82 & 0.82 & 0.81 & 0.69 & 0.75 \\ 
  &  & 12.00 & 0.79 & 0.79 & 0.78 & 0.65 & 0.71 \\ 
  &  & 15.00 & 0.81 & 0.80 & 0.80 & 0.68 & 0.74 \\ 
  &  & 18.00 & 0.85 & 0.85 & 0.85 & 0.73 & 0.81 \\ 
  &  & 21.00 & 0.90 & 0.90 & 0.90 & 0.82 & 0.89 \\ 
  \textcolor{saale}{\textbf{franconian}}& \textbf{tempertature} & 24.00 & 0.93 & 0.92 & 0.93 & 0.87 & 0.93 \\ \cline{2-8}
  & \textbf{radiation}  & 3.00 & 0.75 & 0.69 & 0.69 & 0.64 & 0.70 \\ 
  &  & 6.00 & 0.22 & 0.14 & 0.15 & 0.08 & 0.14 \\ 
  &  & 9.00 & -0.23 & -0.24 & -0.23 & -0.25 & -0.25 \\ 
  &  & 12.00 & -0.39 & -0.36 & -0.35 & -0.36 & -0.38 \\ 
  &  & 15.00 & -0.23 & -0.25 & -0.24 & -0.25 & -0.26 \\ 
  &  & 18.00 & 0.19 & 0.12 & 0.12 & 0.07 & 0.12 \\ 
  &  & 21.00 & 0.69 & 0.63 & 0.63 & 0.59 & 0.63 \\ 
  &  & 24.00 & 0.92 & 0.91 & 0.91 & 0.91 & 0.89 \\ 
  \hline
  &  & 3.00 & 0.96 & 0.96 & 0.95 & 0.93 & 0.97 \\ 
  &  & 6.00 & 0.89 & 0.88 & 0.87 & 0.84 & 0.91 \\ 
  &  & 9.00 & 0.83 & 0.81 & 0.80 & 0.77 & 0.85 \\ 
  &  & 12.00 & 0.80 & 0.77 & 0.77 & 0.75 & 0.83 \\ 
  &  & 15.00 & 0.80 & 0.79 & 0.78 & 0.75 & 0.83 \\ 
  &  & 18.00 & 0.84 & 0.83 & 0.82 & 0.80 & 0.86 \\ 
  &  & 21.00 & 0.88 & 0.88 & 0.88 & 0.86 & 0.90 \\ 
  \textcolor{iller}{\textbf{pre-alpine}} & \textbf{tempertature}  & 24.00 & 0.90 & 0.89 & 0.90 & 0.90 & 0.91 \\ \cline{2-8}
  & \textbf{radiation} & 3.00 & 0.72 & 0.66 & 0.67 & 0.63 & 0.68 \\ 
  &  & 6.00 & 0.15 & 0.08 & 0.11 & 0.05 & 0.10 \\ 
  &  & 9.00 & -0.30 & -0.29 & -0.26 & -0.29 & -0.29 \\ 
  &  & 12.00 & -0.45 & -0.39 & -0.38 & -0.40 & -0.40 \\ 
  &  & 15.00 & -0.31 & -0.29 & -0.27 & -0.30 & -0.29 \\ 
  &  & 18.00 & 0.11 & 0.05 & 0.07 & 0.03 & 0.07 \\ 
  &  & 21.00 & 0.63 & 0.57 & 0.57 & 0.57 & 0.58 \\ 
  &  & 24.00 & 0.86 & 0.86 & 0.86 & 0.89 & 0.85 \\ 
  \hline
  &  & 3.00 & 0.98 & 0.96 & 0.96 & 0.91 & 0.95 \\ 
  &  & 6.00 & 0.93 & 0.90 & 0.88 & 0.80 & 0.85 \\ 
  &  & 9.00 & 0.88 & 0.83 & 0.82 & 0.72 & 0.76 \\ 
  &  & 12.00 & 0.86 & 0.79 & 0.79 & 0.70 & 0.73 \\ 
  &  & 15.00 & 0.86 & 0.81 & 0.80 & 0.70 & 0.75 \\ 
  &  & 18.00 & 0.88 & 0.85 & 0.84 & 0.76 & 0.81 \\ 
  &  & 21.00 & 0.90 & 0.89 & 0.89 & 0.85 & 0.89 \\ 
  \textcolor{ziller}{\textbf{alpine}}& \textbf{tempertature} & 24.00 & 0.90 & 0.90 & 0.91 & 0.91 & 0.92 \\ \cline{2-8}
  & \textbf{radiation} & 3.00 & 0.72 & 0.63 & 0.65 & 0.59 & 0.64 \\ 
  &  & 6.00 & 0.13 & 0.04 & 0.09 & -0.01 & 0.03 \\ 
  &  & 9.00 & -0.33 & -0.31 & -0.27 & -0.34 & -0.33 \\ 
  &  & 12.00 & -0.48 & -0.41 & -0.40 & -0.45 & -0.42 \\ 
  &  & 15.00 & -0.34 & -0.31 & -0.28 & -0.35 & -0.33 \\ 
  &  & 18.00 & 0.09 & 0.01 & 0.05 & -0.02 & 0.01 \\ 
  &  & 21.00 & 0.62 & 0.54 & 0.56 & 0.55 & 0.57 \\ 
  &  & 24.00 & 0.87 & 0.87 & 0.86 & 0.91 & 0.87 \\ 
   \hline
\end{tabular}
\caption{Mean lag statistic by catchment for temperature and Precipitation before and after correction and for reference.
This table incorporates information about the mean from \cref{fig:lag_catch}.} 
\label{tab:lag2catch}
\end{table}

\newpage
%Bibliography
\bibliographystyle{abbrvnat}
\bibliography{VBCArxiv}

\begin{thebibliography}{83}
\providecommand{\natexlab}[1]{#1}
\providecommand{\url}[1]{\texttt{#1}}
\expandafter\ifx\csname urlstyle\endcsname\relax
  \providecommand{\doi}[1]{doi: #1}\else
  \providecommand{\doi}{doi: \begingroup \urlstyle{rm}\Url}\fi

\bibitem[Ahmed et~al.(2013)Ahmed, Wang, Silander, Wilson, Allen, Horton, and Anyah]{ahmed2013}
K.~F. Ahmed, G.~Wang, J.~Silander, A.~M. Wilson, J.~M. Allen, R.~Horton, and R.~Anyah.
\newblock Statistical downscaling and bias correction of climate model outputs for climate change impact assessment in the us northeast.
\newblock \emph{Global and Planetary Change}, 100:\penalty0 320--332, 2013.

\bibitem[Allen et~al.(2017)Allen, McAleer, and Singh]{allen2017}
D.~E. Allen, M.~McAleer, and A.~K. Singh.
\newblock Risk measurement and risk modelling using applications of vine copulas.
\newblock \emph{Sustainability}, 9\penalty0 (10):\penalty0 1762, 2017.

\bibitem[Bacig{\'a}l et~al.(2011)Bacig{\'a}l, J{\'a}gr, and Mesiar]{bacigal2011}
T.~Bacig{\'a}l, V.~J{\'a}gr, and R.~Mesiar.
\newblock Non-exchangeable random variables, archimax copulas and their fitting to real data.
\newblock \emph{Kybernetika}, 47\penalty0 (4):\penalty0 519--531, 2011.

\bibitem[Bedford and Cooke(2001)]{bedford2001}
T.~Bedford and R.~M. Cooke.
\newblock Probability density decomposition for conditionally dependent random variables modeled by vines.
\newblock \emph{Annals of Mathematics and Artificial intelligence}, 32:\penalty0 245--268, 2001.

\bibitem[Bedford and Cooke(2002)]{bedford2002}
T.~Bedford and R.~M. Cooke.
\newblock Vines--a new graphical model for dependent random variables.
\newblock \emph{The Annals of Statistics}, 30\penalty0 (4):\penalty0 1031--1068, 2002.

\bibitem[B{\"o}hning et~al.(1999)B{\"o}hning, Dietz, Schlattmann, Mendonca, and Kirchner]{bohning1999}
D.~B{\"o}hning, E.~Dietz, P.~Schlattmann, L.~Mendonca, and U.~Kirchner.
\newblock The zero-inflated poisson model and the decayed, missing and filled teeth index in dental epidemiology.
\newblock \emph{Journal of the Royal Statistical Society Series A: Statistics in Society}, 162\penalty0 (2):\penalty0 195--209, 1999.

\bibitem[Brockwell(2007)]{brockwell2007}
A.~Brockwell.
\newblock Universal residuals: A multivariate transformation.
\newblock \emph{Statistics \& probability letters}, 77\penalty0 (14):\penalty0 1473--1478, 2007.

\bibitem[Bárdossy and Pegram(2012)]{bardossy2012}
A.~Bárdossy and G.~Pegram.
\newblock Multiscale spatial recorrelation of rcm precipitation to produce unbiased climate change scenarios over large areas and small.
\newblock \emph{Water Resources Research}, 48\penalty0 (9), 2012.
\newblock \doi{https://doi.org/10.1029/2011WR011524}.
\newblock URL \url{https://agupubs.onlinelibrary.wiley.com/doi/abs/10.1029/2011WR011524}.

\bibitem[Cannon(2018)]{cannon2018}
A.~J. Cannon.
\newblock Multivariate quantile mapping bias correction: an n-dimensional probability density function transform for climate model simulations of multiple variables.
\newblock \emph{Climate dynamics}, 50:\penalty0 31--49, 2018.

\bibitem[Cannon(2023)]{cannon2023}
A.~J. Cannon.
\newblock \emph{MBC: Multivariate Bias Correction of Climate Model Outputs}, 2023.
\newblock URL \url{https://CRAN.R-project.org/package=MBC}.
\newblock R package version 0.10-6.

\bibitem[Cannon et~al.(2015)Cannon, Sobie, and Murdock]{cannon2015}
A.~J. Cannon, S.~R. Sobie, and T.~Q. Murdock.
\newblock Bias correction of gcm precipitation by quantile mapping: how well do methods preserve changes in quantiles and extremes?
\newblock \emph{Journal of Climate}, 28\penalty0 (17):\penalty0 6938--6959, 2015.

\bibitem[Charpentier et~al.(2014)Charpentier, Fougères, Genest, and Nešlehová]{charpentier2014}
A.~Charpentier, A.-L. Fougères, C.~Genest, and J.~Nešlehová.
\newblock Multivariate archimax copulas.
\newblock \emph{Journal of Multivariate Analysis}, 126:\penalty0 118--136, 2014.
\newblock ISSN 0047-259X.
\newblock \doi{https://doi.org/10.1016/j.jmva.2013.12.013}.
\newblock URL \url{https://www.sciencedirect.com/science/article/pii/S0047259X14000074}.

\bibitem[Chen et~al.(2013)Chen, Brissette, Chaumont, and Braun]{chen2013}
J.~Chen, F.~P. Brissette, D.~Chaumont, and M.~Braun.
\newblock Finding appropriate bias correction methods in downscaling precipitation for hydrologic impact studies over north america.
\newblock \emph{Water Resources Research}, 49\penalty0 (7):\penalty0 4187--4205, 2013.

\bibitem[Chen et~al.(2021)Chen, Arsenault, Brissette, and Zhang]{chen2021}
J.~Chen, R.~Arsenault, F.~P. Brissette, and S.~Zhang.
\newblock Climate change impact studies: Should we bias correct climate model outputs or post-process impact model outputs?
\newblock \emph{Water Resources Research}, 57\penalty0 (5):\penalty0 e2020WR028638, 2021.

\bibitem[Czado(2019)]{czado2019}
C.~Czado.
\newblock Analyzing dependent data with vine copulas.
\newblock \emph{Lecture Notes in Statistics, Springer}, 2019.

\bibitem[Czado and Nagler(2022)]{czado2022}
C.~Czado and T.~Nagler.
\newblock Vine copula based modeling.
\newblock \emph{Annual Review of Statistics and Its Application}, 9:\penalty0 453--477, 2022.

\bibitem[Dekens et~al.(2017)Dekens, Parey, Grandjacques, and Dacunha-Castelle]{dekens2017}
L.~Dekens, S.~Parey, M.~Grandjacques, and D.~Dacunha-Castelle.
\newblock Multivariate distribution correction of climate model outputs: A generalization of quantile mapping approaches.
\newblock \emph{Environmetrics}, 28\penalty0 (6):\penalty0 e2454, 2017.

\bibitem[Dissmann et~al.(2013)Dissmann, Brechmann, Czado, and Kurowicka]{dissmann2013}
J.~Dissmann, E.~C. Brechmann, C.~Czado, and D.~Kurowicka.
\newblock Selecting and estimating regular vine copulae and application to financial returns.
\newblock \emph{Computational Statistics \& Data Analysis}, 59:\penalty0 52--69, 2013.

\bibitem[Du et~al.(2015)Du, Xiong, Xu, Gippel, Guo, and Liu]{du2015}
T.~Du, L.~Xiong, C.-Y. Xu, C.~J. Gippel, S.~Guo, and P.~Liu.
\newblock Return period and risk analysis of nonstationary low-flow series under climate change.
\newblock \emph{Journal of Hydrology}, 527:\penalty0 234--250, 2015.

\bibitem[Emami and Koch(2018)]{emami2018}
F.~Emami and M.~Koch.
\newblock Evaluation of statistical-downscaling/bias-correction methods to predict hydrologic responses to climate change in the zarrine river basin, iran.
\newblock \emph{Climate}, 6\penalty0 (2):\penalty0 30, 2018.

\bibitem[Erhardt et~al.(2015)Erhardt, Czado, and Schepsmeier]{erhardt2015}
T.~M. Erhardt, C.~Czado, and U.~Schepsmeier.
\newblock R-vine models for spatial time series with an application to daily mean temperature.
\newblock \emph{Biometrics}, 71\penalty0 (2):\penalty0 323--332, 2015.

\bibitem[Fang et~al.(2015)Fang, Yang, Chen, and Zammit]{fang2015}
G.~Fang, J.~Yang, Y.~Chen, and C.~Zammit.
\newblock Comparing bias correction methods in downscaling meteorological variables for a hydrologic impact study in an arid area in china.
\newblock \emph{Hydrology and Earth System Sciences}, 19\penalty0 (6):\penalty0 2547--2559, 2015.

\bibitem[Fran{\c{c}}ois et~al.(2020)Fran{\c{c}}ois, Vrac, Cannon, Robin, and Allard]{franccois2020}
B.~Fran{\c{c}}ois, M.~Vrac, A.~J. Cannon, Y.~Robin, and D.~Allard.
\newblock Multivariate bias corrections of climate simulations: which benefits for which losses?
\newblock \emph{Earth System Dynamics}, 11\penalty0 (2):\penalty0 537--562, 2020.

\bibitem[Funk(2024{\natexlab{a}})]{funk2024a}
H.~Funk.
\newblock {Bias Correction of CRCM5-LE for Hydrological Bavaria}, Oct. 2024{\natexlab{a}}.
\newblock URL \url{https://doi.org/10.5281/zenodo.13348397}.

\bibitem[Funk(2024{\natexlab{b}})]{funk2024b}
H.~Funk.
\newblock \emph{VBC: Vine Copula based Bias Correction for Climate Models}, 2024{\natexlab{b}}.
\newblock URL \url{https://henrifnk.github.io/VBC/}.
\newblock R package version 0.0.1.0.

\bibitem[Geenens and Wang(2018)]{geenens2018}
G.~Geenens and C.~Wang.
\newblock Local-likelihood transformation kernel density estimation for positive random variables.
\newblock \emph{Journal of Computational and Graphical Statistics}, 27\penalty0 (4):\penalty0 822--835, 2018.

\bibitem[Geenens et~al.(2017)Geenens, Charpentier, and Paindaveine]{geenens2017}
G.~Geenens, A.~Charpentier, and D.~Paindaveine.
\newblock {Probit transformation for nonparametric kernel estimation of the copula density}.
\newblock \emph{Bernoulli}, 23\penalty0 (3):\penalty0 1848 -- 1873, 2017.
\newblock \doi{10.3150/15-BEJ798}.
\newblock URL \url{https://doi.org/10.3150/15-BEJ798}.

\bibitem[Guo et~al.(2020)Guo, Chen, Zhang, Xu, and Chen]{guo2020}
Q.~Guo, J.~Chen, X.~J. Zhang, C.-Y. Xu, and H.~Chen.
\newblock Impacts of using state-of-the-art multivariate bias correction methods on hydrological modeling over north america.
\newblock \emph{Water Resources Research}, 56\penalty0 (5):\penalty0 e2019WR026659, 2020.

\bibitem[Haerter et~al.(2011)Haerter, Hagemann, Moseley, and Piani]{haerter2011}
J.~Haerter, S.~Hagemann, C.~Moseley, and C.~Piani.
\newblock Climate model bias correction and the role of timescales.
\newblock \emph{Hydrology and Earth System Sciences}, 15\penalty0 (3):\penalty0 1065--1079, 2011.

\bibitem[Haff et~al.(2010)Haff, Aas, and Frigessi]{haff2010}
I.~H. Haff, K.~Aas, and A.~Frigessi.
\newblock On the simplified pair-copula construction—simply useful or too simplistic?
\newblock \emph{Journal of Multivariate Analysis}, 101\penalty0 (5):\penalty0 1296--1310, 2010.

\bibitem[Hashino et~al.(2007)Hashino, Bradley, and Schwartz]{hashino2007}
T.~Hashino, A.~Bradley, and S.~Schwartz.
\newblock Evaluation of bias-correction methods for ensemble streamflow volume forecasts.
\newblock \emph{Hydrology and Earth System Sciences}, 11\penalty0 (2):\penalty0 939--950, 2007.

\bibitem[Joe(2014)]{joe2014}
H.~Joe.
\newblock \emph{Dependence modeling with copulas}.
\newblock CRC press, 2014.

\bibitem[Kiriliouk et~al.(2023)Kiriliouk, Lee, and Segers]{kiriliouk2023}
A.~Kiriliouk, J.~Lee, and J.~Segers.
\newblock X-vine models for multivariate extremes.
\newblock \emph{arXiv preprint arXiv:2312.15205}, 2023.

\bibitem[Lambert(1992)]{lambert1992}
D.~Lambert.
\newblock Zero-inflated poisson regression, with an application to defects in manufacturing.
\newblock \emph{Technometrics}, pages 1--14, 1992.

\bibitem[Leduc et~al.(2019)Leduc, Mailhot, Frigon, Martel, Ludwig, Brietzke, Gigu{\`e}re, Brissette, Turcotte, Braun, et~al.]{leduc2019}
M.~Leduc, A.~Mailhot, A.~Frigon, J.-L. Martel, R.~Ludwig, G.~B. Brietzke, M.~Gigu{\`e}re, F.~Brissette, R.~Turcotte, M.~Braun, et~al.
\newblock The climex project: A 50-member ensemble of climate change projections at 12-km resolution over europe and northeastern north america with the canadian regional climate model (crcm5).
\newblock \emph{Journal of Applied Meteorology and Climatology}, 58\penalty0 (4):\penalty0 663--693, 2019.

\bibitem[Loader(2006)]{loader2006}
C.~Loader.
\newblock \emph{Local regression and likelihood}.
\newblock Springer Science \& Business Media, 2006.

\bibitem[Maity et~al.(2019)Maity, Suman, Laux, and Kunstmann]{maity2019}
R.~Maity, M.~Suman, P.~Laux, and H.~Kunstmann.
\newblock Bias correction of zero-inflated rcm precipitation fields: a copula-based scheme for both mean and extreme conditions.
\newblock \emph{Journal of Hydrometeorology}, 20\penalty0 (4):\penalty0 595--611, 2019.

\bibitem[Martynov et~al.(2013)Martynov, Laprise, Sushama, Winger, {\v{S}}eparovi{\'c}, and Dugas]{martynov2013}
A.~Martynov, R.~Laprise, L.~Sushama, K.~Winger, L.~{\v{S}}eparovi{\'c}, and B.~Dugas.
\newblock Reanalysis-driven climate simulation over cordex north america domain using the canadian regional climate model, version 5: model performance evaluation.
\newblock \emph{Climate Dynamics}, 41:\penalty0 2973--3005, 2013.

\bibitem[McNeil et~al.(2015)McNeil, Frey, and Embrechts]{mcneil2015}
A.~J. McNeil, R.~Frey, and P.~Embrechts.
\newblock \emph{Quantitative risk management: concepts, techniques and tools-revised edition}.
\newblock Princeton university press, 2015.

\bibitem[Mehrotra and Sharma(2016)]{mehrotra2016}
R.~Mehrotra and A.~Sharma.
\newblock A multivariate quantile-matching bias correction approach with auto-and cross-dependence across multiple time scales: Implications for downscaling.
\newblock \emph{Journal of Climate}, 29\penalty0 (10):\penalty0 3519--3539, 2016.

\bibitem[Mehrotra and Sharma(2021)]{mehrotra2021}
R.~Mehrotra and A.~Sharma.
\newblock A robust alternative for correcting systematic biases in multi-variable climate model simulations.
\newblock \emph{Environmental Modelling \& Software}, 139:\penalty0 105019, 2021.

\bibitem[Montes-Iturrizaga and Heredia-Zavoni(2015)]{montes2015}
R.~Montes-Iturrizaga and E.~Heredia-Zavoni.
\newblock Environmental contours using copulas.
\newblock \emph{Applied Ocean Research}, 52:\penalty0 125--139, 2015.

\bibitem[Morales-Nápoles(2011)]{morales2011}
O.~Morales-Nápoles.
\newblock \emph{Dependence Modeling: Vine Copula Handbook}, chapter Counting vines, page 189–218.
\newblock D Kurowicka, H Joe. World Scientific Publishing Co., 2011.

\bibitem[Mpelasoka and Chiew(2009)]{mpelasoka2009}
F.~S. Mpelasoka and F.~H. Chiew.
\newblock Influence of rainfall scenario construction methods on runoff projections.
\newblock \emph{Journal of Hydrometeorology}, 10\penalty0 (5):\penalty0 1168--1183, 2009.

\bibitem[Nagler(2014)]{nagler2014}
T.~Nagler.
\newblock \emph{Kernel methods for vine copula estimation}.
\newblock Master Thesis, Technische Universität München, 2014.

\bibitem[Nagler(2024)]{nagler2024b}
T.~Nagler.
\newblock Simplified vine copula models: state of science and affairs.
\newblock \emph{arXiv preprint arXiv:2410.16806}, 2024.

\bibitem[Nagler and Vatter(2017)]{nagler2017}
T.~Nagler and T.~Vatter.
\newblock vinecopulib: High performance algorithms for vine copula modeling in c++, 2017.

\bibitem[Nagler and Vatter(2024)]{nagler2024}
T.~Nagler and T.~Vatter.
\newblock Package ‘kde1d’, 2024.

\bibitem[Nahar et~al.(2018)Nahar, Johnson, and Sharma]{nahar2018}
J.~Nahar, F.~Johnson, and A.~Sharma.
\newblock Addressing spatial dependence bias in climate model simulations—an independent component analysis approach.
\newblock \emph{Water Resources Research}, 54\penalty0 (2):\penalty0 827--841, 2018.

\bibitem[Patton(2006)]{patton2006}
A.~J. Patton.
\newblock Modelling asymmetric exchange rate dependence.
\newblock \emph{International economic review}, 47\penalty0 (2):\penalty0 527--556, 2006.

\bibitem[Piani and Haerter(2012)]{piani2012}
C.~Piani and J.~Haerter.
\newblock Two dimensional bias correction of temperature and precipitation copulas in climate models.
\newblock \emph{Geophysical Research Letters}, 39\penalty0 (20), 2012.

\bibitem[Pitie et~al.(2005)Pitie, Kokaram, and Dahyot]{pitie2005}
F.~Pitie, A.~C. Kokaram, and R.~Dahyot.
\newblock N-dimensional probability density function transfer and its application to color transfer.
\newblock In \emph{Tenth IEEE International Conference on Computer Vision (ICCV'05) Volume 1}, volume~2, pages 1434--1439. IEEE, 2005.

\bibitem[Preisser et~al.(2012)Preisser, Stamm, Long, and Kincade]{preisser2012}
J.~S. Preisser, J.~W. Stamm, D.~L. Long, and M.~E. Kincade.
\newblock Review and recommendations for zero-inflated count regression modeling of dental caries indices in epidemiological studies.
\newblock \emph{Caries research}, 46\penalty0 (4):\penalty0 413--423, 2012.

\bibitem[Pörtner et~al.(2022)Pörtner, Roberts, Adams, Adelekan, Adler, Adrian, Aldunce, Ali, Begum, Friedl, Kerr, Biesbroek, Birkmann, Bowen, Caretta, Carnicer, Castellanos, Cheong, Chow, G.~Cissé, and Ibrahim]{ipcc2022}
H.-O. Pörtner, D.~Roberts, H.~Adams, I.~Adelekan, C.~Adler, R.~Adrian, P.~Aldunce, E.~Ali, R.~A. Begum, B.~B. Friedl, R.~B. Kerr, R.~Biesbroek, J.~Birkmann, K.~Bowen, M.~Caretta, J.~Carnicer, E.~Castellanos, T.~Cheong, W.~Chow, G.~C. G.~Cissé, and Z.~Z. Ibrahim.
\newblock \emph{Climate Change 2022: Impacts, Adaptation and Vulnerability}.
\newblock Technical Summary. Cambridge University Press, Cambridge, UK and New York, USA, 2022.
\newblock ISBN 9781009325844.

\bibitem[Reiter et~al.(2016)Reiter, Gutjahr, Schefczyk, Heinemann, and Casper]{reiter2016}
P.~Reiter, O.~Gutjahr, L.~Schefczyk, G.~Heinemann, and M.~Casper.
\newblock Bias correction of ensembles precipitation data with focus on the effect of the length of the calibration period.
\newblock \emph{Meteorologische Zeitschrift}, 25:\penalty0 85--96, 2016.

\bibitem[Robin et~al.(2019)Robin, Vrac, Naveau, and Yiou]{robin2019}
Y.~Robin, M.~Vrac, P.~Naveau, and P.~Yiou.
\newblock Multivariate stochastic bias corrections with optimal transport.
\newblock \emph{Hydrology and Earth System Sciences}, 23\penalty0 (2):\penalty0 773--786, 2019.

\bibitem[Rosenblatt(1952)]{rosenblatt1952}
M.~Rosenblatt.
\newblock Remarks on a multivariate transformation.
\newblock \emph{The annals of mathematical statistics}, 23\penalty0 (3):\penalty0 470--472, 1952.

\bibitem[Salvadori et~al.(2007)Salvadori, De~Michele, Kottegoda, and Rosso]{salvadori2007}
G.~Salvadori, C.~De~Michele, N.~T. Kottegoda, and R.~Rosso.
\newblock \emph{Extremes in nature: an approach using copulas}, volume~56.
\newblock Springer Science \& Business Media, 2007.

\bibitem[{\v{S}}eparovi{\'c} et~al.(2013){\v{S}}eparovi{\'c}, Alexandru, Laprise, Martynov, Sushama, Winger, Tete, and Valin]{separovic2013}
L.~{\v{S}}eparovi{\'c}, A.~Alexandru, R.~Laprise, A.~Martynov, L.~Sushama, K.~Winger, K.~Tete, and M.~Valin.
\newblock Present climate and climate change over north america as simulated by the fifth-generation canadian regional climate model.
\newblock \emph{Climate Dynamics}, 41\penalty0 (11-12):\penalty0 3167--3201, 2013.

\bibitem[Sheather and Jones(1991)]{sheather1991}
S.~J. Sheather and M.~C. Jones.
\newblock A reliable data-based bandwidth selection method for kernel density estimation.
\newblock \emph{Journal of the Royal Statistical Society: Series B (Methodological)}, 53\penalty0 (3):\penalty0 683--690, 1991.

\bibitem[Shi and Yang(2018)]{shi2018}
P.~Shi and L.~Yang.
\newblock Pair copula constructions for insurance experience rating.
\newblock \emph{Journal of the American Statistical Association}, 113\penalty0 (521):\penalty0 122--133, 2018.
\newblock \doi{10.1080/01621459.2017.1330692}.

\bibitem[Sklar(1959)]{sklar1959}
M.~Sklar.
\newblock Fonctions de r{\'e}partition {\`a} n dimensions et leurs marges.
\newblock \emph{Annales de l'ISUP}, 8\penalty0 (3):\penalty0 229--231, 1959.

\bibitem[Soriano et~al.(2019)Soriano, Mediero, and Garijo]{soriano2019}
E.~Soriano, L.~Mediero, and C.~Garijo.
\newblock Selection of bias correction methods to assess the impact of climate change on flood frequency curves.
\newblock \emph{Water}, 11\penalty0 (11):\penalty0 2266, 2019.

\bibitem[S{\o}rland et~al.(2018)S{\o}rland, Sch{\"a}r, L{\"u}thi, and Kjellstr{\"o}m]{sorland2018}
S.~L. S{\o}rland, C.~Sch{\"a}r, D.~L{\"u}thi, and E.~Kjellstr{\"o}m.
\newblock Bias patterns and climate change signals in gcm-rcm model chains.
\newblock \emph{Environmental Research Letters}, 13\penalty0 (7):\penalty0 074017, 2018.

\bibitem[Srikanthan et~al.(2006)Srikanthan, Sharma, and McMahon]{srikanthan2006}
R.~Srikanthan, A.~Sharma, and T.~McMahon.
\newblock Comparison of two nonparametric alternatives for stochastic generation of monthly rainfall.
\newblock \emph{Journal of Hydrologic Engineering}, 11\penalty0 (3):\penalty0 222--229, 2006.

\bibitem[St\"ober(2013)]{stoeber2013b}
J.~St\"ober.
\newblock \emph{Regular Vine Copulas with the simplifying assumption, time-variation, and mixed discrete and continuous margins}.
\newblock PhD thesis, Technische Universität München, 2013.
\newblock URL \url{https://mediatum.ub.tum.de/1137287}.

\bibitem[Stoeber et~al.(2013)Stoeber, Joe, and Czado]{stoeber2013a}
J.~Stoeber, H.~Joe, and C.~Czado.
\newblock Simplified pair copula constructions—limitations and extensions.
\newblock \emph{Journal of Multivariate Analysis}, 119:\penalty0 101--118, 2013.

\bibitem[Sun et~al.(2021)Sun, Huang, Fan, Zhou, Lu, and Wang]{sun2021}
C.~Sun, G.~Huang, Y.~Fan, X.~Zhou, C.~Lu, and X.~Wang.
\newblock Vine copula ensemble downscaling for precipitation projection over the loess plateau based on high-resolution multi-rcm outputs.
\newblock \emph{Water Resources Research}, 57\penalty0 (1):\penalty0 2020WR027698, 2021.

\bibitem[Vaittinada~Ayar et~al.(2021)Vaittinada~Ayar, Vrac, and Mailhot]{vaittinada2021}
P.~Vaittinada~Ayar, M.~Vrac, and A.~Mailhot.
\newblock Ensemble bias correction of climate simulations: preserving internal variability.
\newblock \emph{Scientific Reports}, 11\penalty0 (1):\penalty0 3098, 2021.

\bibitem[Vicente-Serrano et~al.(2021)Vicente-Serrano, Dom{\'\i}nguez-Castro, Murphy, Hannaford, Reig, Pe{\~n}a-Angulo, Tramblay, Trigo, Mac~Donald, Luna, et~al.]{vicente2021}
S.~M. Vicente-Serrano, F.~Dom{\'\i}nguez-Castro, C.~Murphy, J.~Hannaford, F.~Reig, D.~Pe{\~n}a-Angulo, Y.~Tramblay, R.~M. Trigo, N.~Mac~Donald, M.~Y. Luna, et~al.
\newblock Long-term variability and trends in meteorological droughts in western europe (1851--2018).
\newblock \emph{International journal of climatology}, 41:\penalty0 E690--E717, 2021.

\bibitem[Villani(2008)]{villani2008}
C.~Villani.
\newblock \emph{Optimal transport: old and new}, volume 338, chapter The Wasserstein distances, pages 95--112.
\newblock Springer Science \& Business Media, 2008.

\bibitem[von Trentini et~al.(2019)von Trentini, Leduc, and Ludwig]{vonTrentini2019}
F.~von Trentini, M.~Leduc, and R.~Ludwig.
\newblock Assessing natural variability in rcm signals: comparison of a multi model euro-cordex ensemble with a 50-member single model large ensemble.
\newblock \emph{Climate Dynamics}, 53:\penalty0 1963--1979, 2019.

\bibitem[Vrac(2018)]{vrac2018}
M.~Vrac.
\newblock Multivariate bias adjustment of high-dimensional climate simulations: the rank resampling for distributions and dependences (r 2 d 2) bias correction.
\newblock \emph{Hydrology and Earth System Sciences}, 22\penalty0 (6):\penalty0 3175--3196, 2018.

\bibitem[Vrac and Friederichs(2015)]{vrac2015}
M.~Vrac and P.~Friederichs.
\newblock Multivariate—intervariable, spatial, and temporal—bias correction.
\newblock \emph{Journal of Climate}, 28\penalty0 (1):\penalty0 218--237, 2015.

\bibitem[Vrac et~al.(2022)Vrac, Thao, and Yiou]{vrac2022}
M.~Vrac, S.~Thao, and P.~Yiou.
\newblock Should multivariate bias corrections of climate simulations account for changes of rank correlation over time?
\newblock \emph{Journal of Geophysical Research: Atmospheres}, 127\penalty0 (14):\penalty0 e2022JD036562, 2022.

\bibitem[Westra et~al.(2013)Westra, Evans, Mehrotra, and Sharma]{westra2013}
S.~Westra, J.~P. Evans, R.~Mehrotra, and A.~Sharma.
\newblock A conditional disaggregation algorithm for generating fine time-scale rainfall data in a warmer climate.
\newblock \emph{Journal of Hydrology}, 479:\penalty0 86--99, 2013.

\bibitem[Willkofer et~al.(2018)Willkofer, Schmid, Komischke, Korck, Braun, and Ludwig]{willkofer2018}
F.~Willkofer, F.-J. Schmid, H.~Komischke, J.~Korck, M.~Braun, and R.~Ludwig.
\newblock The impact of bias correcting regional climate model results on hydrological indicators for bavarian catchments.
\newblock \emph{Journal of Hydrology: Regional Studies}, 19:\penalty0 25--41, 2018.

\bibitem[Willkofer et~al.(2020)Willkofer, Wood, von Trentini, Weism{\"u}ller, Poschlod, and Ludwig]{willkofer2020}
F.~Willkofer, R.~R. Wood, F.~von Trentini, J.~Weism{\"u}ller, B.~Poschlod, and R.~Ludwig.
\newblock A holistic modelling approach for the estimation of return levels of peak flows in bavaria.
\newblock \emph{Water}, 12\penalty0 (9):\penalty0 2349, 2020.

\bibitem[Wood(2024)]{wood2024}
R.~R. Wood.
\newblock Sdclirefv2, Aug. 2024.
\newblock URL \url{https://doi.org/10.5281/zenodo.13221576}.

\bibitem[Wood et~al.(2017)Wood, Willkofer, Schmid, Trentini, Komischke, and Ludwig]{wood2017}
R.~R. Wood, F.~Willkofer, F.-J. Schmid, F.~Trentini, H.~Komischke, and R.~Ludwig.
\newblock Sdcliref-a sub-daily gridded reference dataset.
\newblock In \emph{EGU General Assembly Conference Abstracts}, page 15739, 2017.

\bibitem[Yip and Yau(2005)]{yip2005}
K.~C. Yip and K.~K. Yau.
\newblock On modeling claim frequency data in general insurance with extra zeros.
\newblock \emph{Insurance: Mathematics and Economics}, 36\penalty0 (2):\penalty0 153--163, 2005.

\bibitem[Zscheischler and Fischer(2020)]{zscheischler2020}
J.~Zscheischler and E.~M. Fischer.
\newblock The record-breaking compound hot and dry 2018 growing season in germany.
\newblock \emph{Weather and Climate Extremes}, 29:\penalty0 100270, 2020.

\bibitem[Zscheischler et~al.(2018)Zscheischler, Westra, Van Den~Hurk, Seneviratne, Ward, Pitman, AghaKouchak, Bresch, Leonard, Wahl, et~al.]{zscheischler2018}
J.~Zscheischler, S.~Westra, B.~J. Van Den~Hurk, S.~I. Seneviratne, P.~J. Ward, A.~Pitman, A.~AghaKouchak, D.~N. Bresch, M.~Leonard, T.~Wahl, et~al.
\newblock Future climate risk from compound events.
\newblock \emph{Nature Climate Change}, 8\penalty0 (6):\penalty0 469--477, 2018.

\end{thebibliography}

\end{document}